\documentclass[12pt, a4paper]{article} 
\baselineskip=\normalbaselineskip

\usepackage{mathrsfs,amsbsy,amssymb,latexsym,amsfonts,amsmath,amsthm}
\usepackage{graphicx,color}

\usepackage{bm}

\newcommand{\bbC}{{\mathbb{C}}}
\newcommand{\bbR}{{\mathbb{R}}}

\newcommand{\calR}{{\mathcal{R}}}
\newcommand{\lapse}{{\alpha}}
\newcommand{\shift}{{\beta}}
\newcommand{\wt}{{W}}
\newcommand{\denom}{{A}}
\newcommand{\muR}{{Dz}}

\newcommand{\muSigmat}{{|dz_t|}}
\newcommand{\muTR}{{D\pi}}
\newcommand{\muTRp}{{D\pi'}}
\newcommand{\idxq}{{c}}

\makeatletter
\catcode`\@=11
\@addtoreset{equation}{section}

\def\@seccntformat#1{\csname the#1\endcsname.~~}
\makeatother

\begin{document}

\begin{titlepage}
\renewcommand{\thefootnote}{\fnsymbol{footnote}}
\begin{flushright}
KUNS-2848
\end{flushright}
\vspace*{1.0cm}

\begin{center}
{\Large \bf
Worldvolume approach to the tempered Lefschetz thimble method
}
\vspace{1.0cm}

\centerline{
{Masafumi Fukuma}%
\footnote{E-mail address: 
fukuma@gauge.scphys.kyoto-u.ac.jp} and
{Nobuyuki Matsumoto}%
\footnote{E-mail address: 
nobu.m@gauge.scphys.kyoto-u.ac.jp}
}

\vskip 0.1cm
{\it Department of Physics, Kyoto University, Kyoto 606-8502, Japan}
\vskip 1.2cm

\end{center}

\begin{abstract}

As a solution towards the numerical sign problem, 
we propose a novel Hybrid Monte Carlo algorithm, 
in which molecular dynamics is performed 
on a continuum set of integration surfaces 
foliated by the antiholomorphic gradient flow
(``the worldvolume of an integration surface''). 
This is an extension of the tempered Lefschetz thimble method (TLTM),
and solves the sign and multimodal problems simultaneously 
as the original TLTM does. 
Furthermore, in this new algorithm, 
one no longer needs to compute 
the Jacobian of the gradient flow in generating a configuration, 
and only needs to evaluate its phase upon measurement. 
To demonstrate that this algorithm works correctly, 
we apply the algorithm to a chiral random matrix model, 
for which the complex Langevin method is known not to work.

\end{abstract}
\end{titlepage}

\pagestyle{empty}
\pagestyle{plain}

\tableofcontents
\setcounter{footnote}{0}

\section{Introduction}
\label{sec:introduction}

The sign problem is one of the major obstacles 
to numerical computation in various areas of physics, 
including 
finite density QCD \cite{Aarts:2015tyj}, 
quantum Monte Carlo simulations of 
statistical systems \cite{Pollet:2012}, 
and the numerical simulations of real-time quantum field theories. 

There have been proposed many Monte Carlo algorithms 
towards solving the sign problem, 
such as those based on the complex Langevin equation 
\cite{Parisi:1984cs,Klauder:1983nn,Klauder:1983sp,
Aarts:2009uq,Aarts:2011ax,Aarts:2013uxa,Nagata:2016vkn}
and those on Lefschetz thimbles 
\cite{Witten:2010cx,Cristoforetti:2012su,Cristoforetti:2013wha,
Fujii:2013sra,Alexandru:2015xva,Alexandru:2015sua,
Fukuma:2017fjq,Alexandru:2017oyw,Alexandru:2017lqr,
Fukuma:2019wbv,Fukuma:2019uot}, 
each of which has its own advantage and disadvantage. 
The advantage to use the Complex Langevin equation 
is its cheap computational cost, 
but such algorithms are known to suffer from a notorious problem 
called the ``wrong convergence problem'' 
(giving incorrect results with small statistical errors) 
for physically important ranges of parameters 
\cite{Aarts:2009uq,Aarts:2011ax,Nagata:2016vkn}. 
On the other hand, 
although computationally expensive, 
the algorithms based on Lefschetz thimbles are 
basically free from the wrong convergence. 
However, this is the case 
when and only when a single Lefschetz thimble is relevant 
for the estimation of observables, 
because otherwise there can appear another problem of multimodality 
due to infinitely high potential barriers 
between different thimbles. 

The tempered Lefschetz thimble method (TLTM) \cite{Fukuma:2017fjq}
is a Lefschetz thimble method 
that avoids the sign and multimodal problems simultaneously, 
by introducing a discrete set of integration surfaces 
(replicas of integration surface)
and exchanging configurations between replicas.  
The TLTM has proved effective and versatile 
when applied to various models, 
including $(0+1)$-dimensional massive Thirring model \cite{Fukuma:2017fjq}, 
the two-dimensional Hubbard model away from half filling.
The disadvantage of the original TLTM is its computational cost, 
the cost coming from the computation of the Jacobian 
and from the additional cost due to the introduction of replicas. 

In this paper, as an extension of the TLTM, 
we propose a novel Hybrid Monte Carlo (HMC) algorithm, 
where molecular dynamics is performed 
on a continuum set of replicas,
not on each replica as was done in \cite{Fukuma:2019uot} 
(see also \cite{Fujii:2013sra,Alexandru:2019}). 
This algorithm no longer requires the computation of the Jacobian 
in generating a configuration, 
which is expensive for large systems. 
To overview this new algorithm, 
we first review the basics of the sign problem, 
and then introduce our algorithm. 

Let us consider a system of an $N$-dimensional dynamical variable 
$x\in\bbR^N$ 
with an action $S(x)$. 
Our aim is to evaluate the expectation value of 
an operator $\mathcal{O}(x)$,
\begin{align}
  \langle \mathcal{O}(x) \rangle
  &\equiv \frac{1}{Z}\,\int_{\bbR^N} d x\, e^{-S(x)} \mathcal{O}(x), 
  \quad Z \equiv \int_{\bbR^N} d x\, e^{-S(x)},
\end{align}
where $dx \equiv dx^1 \cdots dx^N$. 
When the action takes complex values, 
the Boltzmann weight $e^{-S(x)}/Z$ 
can no longer be interpreted as a probability distribution, 
which invalidates a direct application 
of the Markov chain Monte Carlo (MCMC) method. 
The simplest workaround is the so-called reweighting method, 
where a positive weight is constructed 
from the real part of the action, 
$e^{-{\rm Re}\, S(x)} / \int_{\bbR^N}dx\,e^{-{\rm Re}\, S(x)}$, 
and $\langle\mathcal{O}(x)\rangle$ is estimated 
by a ratio of reweighted averages,
\begin{align}
  \langle \mathcal{O}(x) \rangle
  = \frac{ \langle e^{-i\, {\rm Im}\,S(x)}\mathcal{O}(x) 
  \rangle_{\rm rewt} } 
  { \langle e^{-i\, {\rm Im}\,S(x)} \rangle_{\rm rewt}},
\label{reweighting1}
\end{align}
where
\begin{align}
  \langle f(x) \rangle_{\rm rewt} \equiv
  \frac{ \int_{\bbR^N} d x\, e^{- {\rm Re}\,S(x)} f(x) }
  { \int_{\bbR^N} d x\, e^{- {\rm Re}\,S(x)} }. 
\label{reweighting2}
\end{align}
For large degrees of freedom (DOF), however, 
the phase factor $e^{-i\, {\rm Im}\,S(x)}$ in reweighted averages 
makes the integrals highly oscillatory, 
so that \eqref{reweighting1} becomes a ratio of 
exponentially small quantities of $e^{-O(N)}$ 
even when the ratio should give a quantity of $O(1)$. 
Since the reweighting is a mathematically equivalent rewriting, 
it should not give any problems 
if one can obtain the values of the reweighted averages precisely 
both in the numerator and the denominator. 
However, in the Monte Carlo calculations, 
they are evaluated separately as sample averages, 
which should be accompanied by statistical errors 
of $O(1/\sqrt{N_{\rm conf}})$, 
where $N_{\rm conf}$ is the size of the sample. 
Thus, for the naive reweighting method, 
the expectation value is estimated in the form 
\begin{align}
  \langle \mathcal{O}(x) \rangle
  \approx \frac{ e^{-O(N)} \pm O( 1/\sqrt{N_{\rm conf}} ) }
  { e^{-O(N)} \pm O( 1/\sqrt{N_{\rm conf}} ) }, 
\label{rewt_estim}
\end{align}
which means that 
we need an exponentially large sample size, 
$N_{\rm conf} = e^{O(N)}$, 
in order to make the statistical errors relatively small 
compared to the exponentially small mean values. 
This enormous computational time makes the MCMC computations impractical. 
This is the sign problem. 

Lefschetz thimble methods are a class of algorithms 
towards solving the sign problem. 
In these methods, we complexify the integration variables, 
$x\in\bbR^N \rightarrow z\in\bbC^N$,  
with the assumption that 
$e^{-S(z)}$ and $e^{-S(z)}\mathcal{O}(z)$ are entire functions over $\bbC^N$. 
Then, from Cauchy's theorem, 
the integrals do not change under continuous deformations 
of the integration surface 
from $\bbR^N$ to $\Sigma \subset \bbC^N$, 
as long as the boundary at infinity ($|x|\to\infty$) 
is kept fixed under the deformations:
\begin{align}
  \langle \mathcal{O}(x) \rangle &= \frac{\int_\Sigma dz\, 
  e^{-S(z)} \mathcal{O}(z)}
  {\int_\Sigma dz\, e^{-S(z)}}, 
\label{integral_Sigma}
\end{align} 
where $dz \equiv dz^1 \cdots dz^N$. 
We expect that the sign problem is reduced 
if we can find an integration surface $\Sigma$ 
on which $e^{-i\,{\rm Im}\, S(z)}$ is almost constant. 

Such surfaces are obtained 
from the following antiholomorphic gradient flow $z_t(x)$ at large flow times: 
\begin{align}
  \frac{dz_t^i}{dt} &= [\partial_i S(z_t)]^\ast,
  \quad
  z^i_{t=0} = x^i. 
\label{flow_zC}
\end{align}
In fact, this flow defines a map from the original integration surface 
$\Sigma_0\equiv\bbR^N$ 
to a real $N$-dimensional submanifold 
$\Sigma_t\equiv\{z_t(x)\,|\,x \in\bbR^N\}$ in $\bbC^N=\bbR^{2N}$: 
\begin{align}
  z_t\,:~\Sigma_0\ni x \mapsto z_t(x)\in\Sigma_t, 
\end{align}
and the flowed surface 
$\Sigma_t$ approaches in the limit $t\to\infty$ 
a union of Lefschetz thimbles, 
on each of which ${\rm Im}\, S(z)$ is constant.%
\footnote{
  Since $(d/dt)\,S(z_t)=|S(z_t)|^2\geq0$, 
  ${\rm Im}\,S(z_t)$ is constant along the flow, 
  and ${\rm Re}\,S(z_t)$ increases except at critical points $z_\sigma$ 
  [at which $\partial_i S(z_\sigma)=0$]. 
  For each critical point $z_\sigma$, 
  the Lefschetz thimble $\mathcal{J}_\sigma$ is defined as 
  $\mathcal{J}_\sigma\equiv
  \{z\in\bbC^N | \lim_{t\to-\infty} z_t(z)=z_\sigma\}$, 
  on which ${\rm Im}\,S(z)$ is constant [$={\rm Im}\,S(z_\sigma)$]. 
 } 
We thus expect that the sign problem is 
substantially remedied on $\Sigma_t$ 
for sufficiently large $t$. 
The expectation value is then expressed in the form%
\footnote{
  The original reweighting 
  [eqs.~\eqref{reweighting1} and \eqref{reweighting2}] 
  corresponds to the $t=0$ case. 
  When only a single Lefschetz thimble is relevant, 
  one can argue that 
  the exponentially small part in the estimation \eqref{rewt_estim} 
  increases as $e^{-e^{-\lambda t} O(N)}$, 
  where $\lambda$ is the minimum singular value 
  of $H(z)=(\partial_i\partial_j S(z))$ at the critical point. 
  We thus expect that the sign problem is removed 
  for the flow time $t\gtrsim O(\ln N)$. 
} 
\begin{align}
  \langle \mathcal{O}(x) \rangle
  = \frac{ \langle 
  e^{-i\, {\rm Im}\,S(z) + i \varphi(z) }\mathcal{O}(z) \rangle_{\Sigma_t} }
  { \langle e^{-i\, {\rm Im}\,S(z) + i \varphi(z) } \rangle_{\Sigma_t} } 
\label{expec_Sigmat1}
\end{align}
with 
\begin{align}
  \langle f(z) \rangle_{\Sigma_t} \equiv
  \frac{ \int_{\Sigma_t} \muSigmat \, e^{- {\rm Re}\,S(z_t)} f(z_t) }
  { \int_{\Sigma_t} \muSigmat \, e^{- {\rm Re}\,S(z_t)} }. 
\label{expec_Sigmat2}
\end{align}
Here, $\muSigmat$ is the invariant volume element of $\Sigma_t$, 
and can be expressed with the Jacobian of the flow, 
$J_t(x) \equiv \partial z_t(x)/\partial x$, as
\begin{align}
  dz_t = \det J_t \, d x,
  \quad 
  |dz_t| = |\det J_t| \, d x. 
\end{align}
The phase factor $e^{i\varphi(z)}$ 
in \eqref{expec_Sigmat1} is then given by 
\begin{align}
  e^{i\varphi(z)} \equiv \frac{dz_t}{|dz_t|} = \frac{\det J_t}{|\det J_t|}. 
\label{measure_varphi}
\end{align}
The Jacobian $J_t(x)$ can be computed 
by solving the second flow equation:%
\footnote{
  This can be shown as 
  $(d/dt)(J_t(x))^i_{~a}=(\partial/\partial t)(\partial z^i_t(x)/\partial x^a)
  =\partial[\partial z^i_t(x)/\partial t]/\partial x^a
  =\partial[\partial_i S(z_t(x))]^\ast/\partial x^a
  =[\partial_i\partial_j S(z_t(x))\,(\partial z^j_t(x)/\partial x^a))]^\ast
  =[H_{ij}(z_t(x))\,(J_t(x))^j_{~a}]^\ast$.
} 
\begin{align}
  \frac{d J_t}{d t} = [H(z_t) J_t]^\ast,
  \quad
  J_{t=0} = 1, 
\label{Jacobian}
\end{align}
where $H(z) \equiv (\partial_i\partial_j S(z) )$. 

When $\Sigma_t$ approaches more than one Lefschetz thimble, 
$\Sigma_t$ gets decomposed into separate regions 
as $t$ increases, 
each region being surrounded by infinitely high potential barriers. 
This causes a multimodal problem 
in MCMC calculations.%
\footnote{
  In the following discussions, 
  we assume that there is no multimodal problem 
  on the original integration surface $\Sigma_{t=0}=\bbR^N$. 
  If this is not the case, 
  we implement an extra algorithm to resolve the multimodality 
  (such as the tempering 
  with respect to the overall coefficient of the action) 
  or make a shift of the starting integration surface 
  from $\bbR^N$. 
} 
The tempered Lefschetz thimble method (TLTM) was proposed 
in \cite{Fukuma:2017fjq} 
to solve this multimodality 
by implementing the parallel tempering algorithm 
\cite{Swendsen1986,Geyer1991,Hukushima1996}
with the flow time $t$ used as the tempering parameter. 
Namely, we prepare a finite set of flow times, $\{t_\alpha\}$,  
and introduce copies (replicas) of 
the corresponding configuration spaces, 
$\{\Sigma_{t_\alpha}\}$.%
\footnote{
  Note that $\Sigma_t$ is not necessarily homeomorphic to $\Sigma_0=\bbR^N$ 
  because we remove zeros of $e^{-S(z)}$ from $\Sigma_t$ 
  (see \cite{Fukuma:2019uot}).
} 
The set $\{t_\alpha\}$ is chosen 
so as to include large enough flow times 
to resolve the sign problem, 
as well as small enough flow times 
to resolve the multimodality.%
\footnote{
  See \cite{Fukuma:2018qgv,Fukuma:2020got}
  for a geometrical optimization of the values $t_\alpha$ 
  based on the distance between configurations introduced 
in \cite{Fukuma:2017wzs}.
} 
Then, in addition to Monte Carlo updates on each $\Sigma_t$, 
we swap configurations between adjacent replicas, 
which enables easy communications 
between configurations around different modes, 
and thus accelerates the relaxation to global equilibrium. 
Thus, the TLTM is an algorithm that solves 
the sign and multimodal problems simultaneously, 
and has proved effective for various models 
\cite{Fukuma:2017fjq,Fukuma:2019wbv,Fukuma:2019uot}, 
as mentioned before. 
However, in this original TLTM, 
we need to increase the number of replicas 
as we increase DOF 
in order to keep sufficient acceptance rates in the swapping process. 
Furthermore, we have to compute the Jacobian $J_t(x)$ 
every time in the swapping process, 
which is expensive 
because the second flow equation \eqref{Jacobian} involves 
a matrix multiplication, 
whose cost is of $O(N^3)$. 

In this paper, 
we propose a novel Hybrid Monte Carlo (HMC) algorithm, 
where molecular dynamics is performed 
on a continuum set of integration surfaces, $\bigcup_t \Sigma_t$.  
This algorithm solves the multimodal problem 
without preparing replicas. 
Furthermore, 
the Jacobian of the gradient flow no longer needs to be computed 
in generating a configuration, 
and only its phase needs to be evaluated upon measurement. 

\begin{figure}[ht]
  \centering
  \includegraphics[width=85mm]{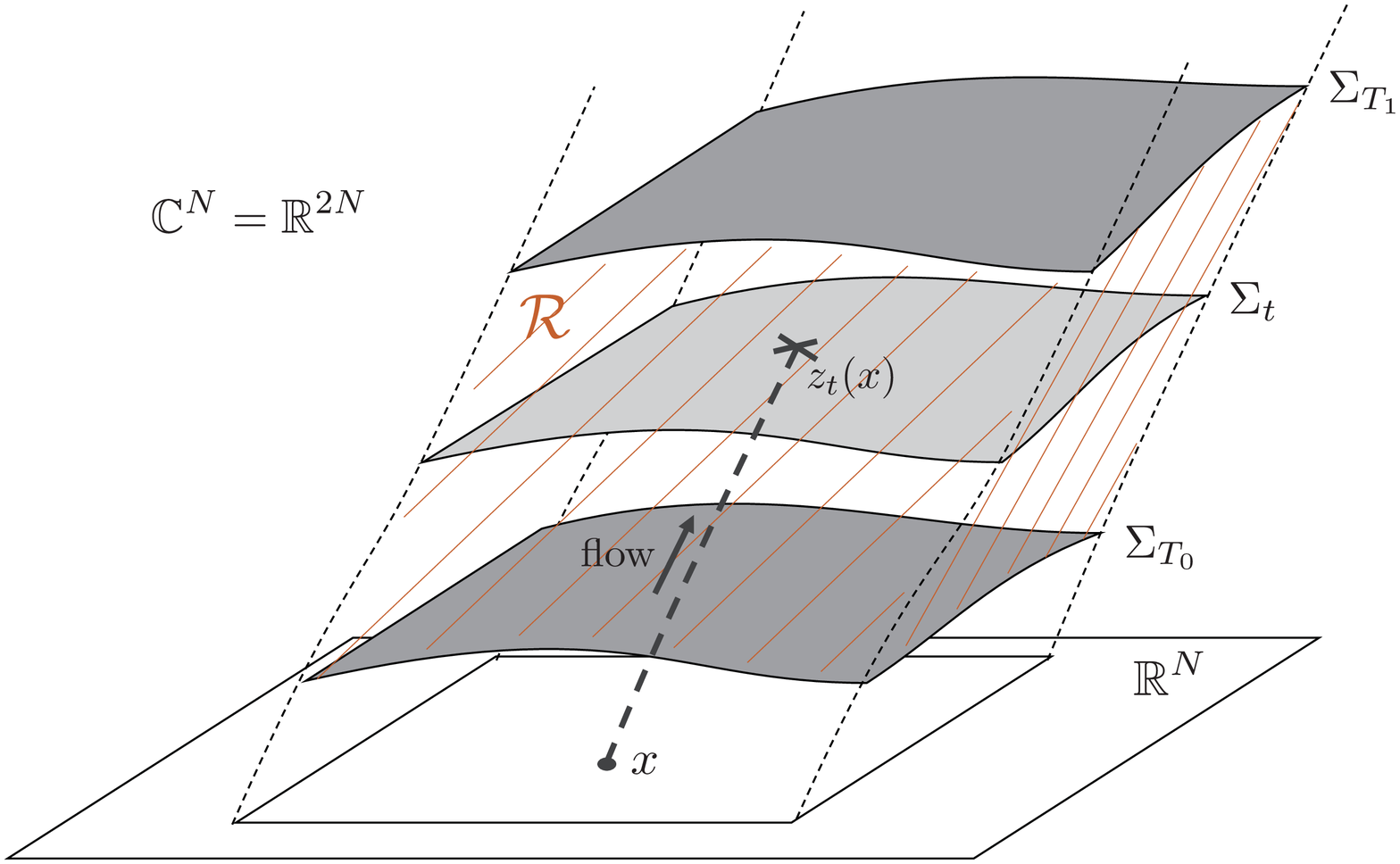} 
  \caption{
  \label{fig:calR}
    The worldvolume $\calR$ (shaded region) embedded in $\bbC^N=\bbR^{2N}$.
  }
\end{figure}

This algorithm is based on the observation that, 
since integrals on $\Sigma_t$ do not depend on $t$ 
due to Cauchy's theorem, 
the values do not change 
even when we average them over $t$ in a range $[T_0,T_1]$ 
with an arbitrary weight 
$e^{- \wt (t)}$: 
\begin{align}
  \langle \mathcal{O}(x) \rangle
  &= \frac{ \int_{T_0}^{T_1} dt\, e^{- \wt(t) }
  \int_{\Sigma_t} dz_t \, e^{-S(z_t)}\, \mathcal{O}(z_t) }
  { \int_{T_0}^{T_1} dt \, e^{- \wt(t)}\int_{\Sigma_t} dz_t \,
  e^{-S(z_t)} } .
\label{integral_R_intro}
\end{align}
We denote the new integration region by $\calR$ 
(see Fig.~\ref{fig:calR}):
\begin{align}
  \calR 
  \equiv\bigcup_{t=T_0}^{T_1}\Sigma_t
  = \bigl\{z_t(x)\in\bbC^N \,\big|\, t \in [T_0,{T_1}],\,x\in{\bbR^N} \bigr\},
\label{worldvolume}
\end{align}
which we regard as the {\em worldvolume} of an integration surface 
moving in the ``target space'' $\bbC^N=\bbR^{2N}$. 
We abbreviate the TLTM based on \eqref{integral_R_intro} 
as the WV-TLTM. 
\noindent 
Although the weight function $e^{-\wt(t)}$ can be chosen arbitrarily, 
a practically good choice is the one 
which gives an almost uniform distribution with respect to $t$ 
(see subsection \ref{sec:weight} for details). 

The expectation value is now expressed 
as a ratio of reweighted averages over $\calR$:
\begin{align}
  \langle \mathcal{O}(x) \rangle
  = \frac{\int_\calR \muR\,e^{-V(z)}\,\denom(z)\,\mathcal{O}(z)}
  {\int_\calR \muR\,e^{-V(z)}\,\denom(z)}
  = \frac{\langle \denom(z)\,\mathcal{O}(z) \rangle_\calR}
  {\langle \denom(z) \rangle_\calR}.
\end{align}
Here, the reweighted average 
\begin{align}
  \langle f(z) \rangle_\calR \equiv 
  \frac{\int_\calR \muR\,e^{-V(z)}\,f(z)}
  {\int_\calR \muR\,e^{-V(z)}}
\label{expec_R}
\end{align}
is defined with respect to the (real-valued) invariant volume element 
$\muR$ on the $(N+1)$-dimensional region $\calR$ 
and to the new weight%
\footnote{
 The function $t=t(z)$ is given by $t$ in $z=z_t(x)$. 
 Later we will extend the defining region from $\calR$ 
 to the vicinity of $\calR$ 
 in order to define the gradient $\partial_i t(z)$ on $\calR$ 
 (see subsection \ref{sec:wv_md}).
} 
\begin{align}
  e^{-V(z)} \equiv e^{-{\rm Re}\,S(z)-\wt(t(z))}.
\label{potential}
\end{align}
The accompanying reweighting factor $\denom(z)$ is then given by 
\begin{align}
  \denom(z) \equiv 
  \frac{ e^{-S(z)-\wt(t(z))} dt\, dz_t }{ e^{-V(z)} \muR } 
  = e^{-i\,{\rm Im}\,S(z)} \frac{dt\, d z_t}{\muR}. 
\label{reweighting_factor}
\end{align}
The aim of this paper is to establish a HMC algorithm 
for the reweighted average \eqref{expec_R} 
on the worldvolume $\calR$. 
To demonstrate that this algorithm works correctly, 
we apply the algorithm to a chiral random matrix model 
(the Stephanov model) 
\cite{Stephanov:1996ki,Halasz:1998qr}, 
for which the complex Langevin method is known 
not to work \cite{Bloch:2017sex}. 

This paper is organized as follows. 
In section \ref{sec:review}, 
we review the basics of the HMC algorithm 
on a general constrained space $\calR$. 
In section \ref{sec:wv_hmc}, 
we deepen the argument 
for the case where the constrained surface $\calR$ 
is the worldvolume of an integration surface. 
We first study the geometry of the worldvolume $\calR$ 
by using the Arnowitt-Deser-Misner parametrization of the metric. 
We then construct molecular dynamics on $\calR$, 
with a prescription to determine the weight $e^{-\wt(t)}$. 
After summarizing the HMC algorithm on the worldvolume $\calR$, 
we give an algorithm to estimate observables.
In section \ref{sec:application}, 
we apply this algorithm to the Stephanov model, 
and show that our algorithm correctly reproduces exact results, 
solving both sign and multimodal problems. 
Section \ref{sec:conclusion} is devoted to conclusion and outlook.

\section{HMC algorithm on a constrained space (review)}
\label{sec:review}

In this section, 
we briefly review the basics of the RATTLE algorithm 
\cite{Andersen:1983,Leimkuhler:1994}, 
which is a HMC algorithm 
on a constrained space such as our worldvolume $\calR$. 
A detailed discussion is given in appendix \ref{sec:RATTLE_geometry} 
with more geometrical terms.

\subsection{Stochastic process on a constrained space}
\label{sec:constrained_stochastic}

Let $\calR$ be an $m$-dimensional manifold 
embedded in the flat space 
$\bbR^M=\{z=(z^I)\}$ $(I=1,\ldots,M)$. 
We assume that $\calR$ is characterized 
by $M-m$ independent constraint equations 
$\phi^r(z)=0$ $(r=1,\ldots,M-m)$. 
When $\calR$ is the worldvolume of an integration surface, 
we set $M=2N$ and $m=N+1$, 
treating $\bbC^N$ as a real space $\bbR^{2N}$. 

Denoting the coordinates on $\calR$ by $\xi=(\xi^\mu)$ $(\mu=1,\ldots,m)$, 
the embedding is expressed by functions $z^I=z^I(\xi)$, 
and the induced metric on $\calR$ is given by 
\begin{align}
  ds^2 &= (dz^I(\xi))^2 \equiv g_{\mu\nu}(\xi)\,d\xi^\mu d\xi^\nu
  ~~\mbox{with}~~
  g_{\mu\nu}(\xi) = \partial_\mu z^I(\xi)\,\partial_\nu z^I(\xi),
\end{align}
which defines the invariant volume element as 
\begin{align}
  \muR \equiv \sqrt{g(\xi)}\,d\xi, 
\end{align}
where $d\xi=d\xi^1\cdots \xi^m$.

The probability distribution $p(z)$ on $\calR$ is defined 
with respect to $\muR$, 
and thus is normalized as $\int_\calR \muR\,p(z) = 1$. 
The transition matrix is also defined for $\muR$, 
so that a transition from a probability distribution $p(z)$ 
to $p'(z)$ ($z\in\calR$) is expressed 
with a transition matrix $P(z'|z)$ as 
\begin{align}
  p'(z') = \int_\calR \muR\,P(z'|z)\,p(z) \quad (z'\in\calR). 
\end{align}
For the equilibrium distribution on $\calR$ 
with respect to a potential $V(z)$, 
\begin{align}
  p_{\rm eq}(z) \equiv e^{-V(z)} / Z_\calR \quad 
  \Bigl(Z_\calR = \int_\calR \muR\,e^{-V(z)} \Bigr),
\end{align}
the detailed balance condition is given by 
\begin{align}
  P(z'|z)\,e^{-V(z)} = P(z|z')\,e^{-V(z')}\quad(z,z'\in\calR).
\end{align}

Throughout this paper, 
we denote a function on $\calR$ by $f(z)$ and $f(\xi)$, 
interchangeably, with the understanding that $z=z(\xi)$.
The transition matrix on $\calR$ is also written 
as $P(z'|z)$ and $P(\xi'|\xi)$ 
for $z=z(\xi),\,z'=z(\xi')\in\calR$.

\subsection{HMC on a constrained space}
\label{sec:constrained_HMC}

Denoting by $\pi=(\pi^I)$ the conjugate momentum to $z=(z^I)\in\calR$, 
we consider the Hamiltonian dynamics on $\calR$ 
with the Hamiltonian 
\begin{align}
  H(z,\pi) = \frac{1}{2}\,(\pi^I)^2 + V(z),
\end{align}
which can be expressed as a set of first-order differential equations 
in time $s$ 
with Lagrange multipliers $\lambda_r$: 
\begin{align}
  \partial_s z &= \pi,
\label{zdot}
\\
  \partial_s \pi &= -\partial V(z) - \lambda_r\, \partial \phi^r(z),
\label{pidot}
\\
  \phi^r(z) &= 0,
\label{constraint_z}
\\
  \pi \cdot \partial \phi^r(z) &= 0.
\label{constraint_pi} 
\end{align}
Here, $\partial \equiv (\partial_{z^I})$ is the gradient in $\bbR^{M}$. 

Equations \eqref{zdot}--\eqref{constraint_pi} 
can be discretized 
such that the symplecticity and the reversibility still hold 
after the discretization (below $\Delta s$ is the step size) 
\cite{Andersen:1983,Leimkuhler:1994}: 
\begin{align}
  \pi_{1/2} &= \pi - \frac{\Delta s}{2}\, \partial V(z) 
  - \lambda_r\, \partial \phi^r(z),
\label{pihalf}
\\
  z' &= z + \Delta s \, \pi_{1/2},
\label{zprime}
\\
  \pi' &= \pi - \frac{\Delta s}{2}\, \partial V(z') 
  - {\lambda'}_r\, \partial \phi^r(z'),
\label{piprime}
\end{align}
where $\lambda_r$ and $\lambda'_r$ are determined, respectively, 
so that the following constraints are satisfied:%
\footnote{
  We regard the tangent bundle $T\calR=\bigcup_z T_z\calR$ 
  ({\em not} the cotangent bundle $T^\ast\calR$) 
  as the phase space for motions on $\calR$ \cite{Andersen:1983,Leimkuhler:1994}. 
  See appendix \ref{sec:RATTLE_geometry} for details. 
} 
\begin{align}
  z' &\in \calR \quad (\mbox{i.e.~}\phi^r(z')=0), 
\label{constraint_zprime}
\\
  \pi' &\in T_{z'} \calR.
\label{constraint_piprime}
\end{align}
One can easily show that 
the map $\Phi_{\Delta s}:\,(z,\pi) \rightarrow (z',\pi')$ 
actually satisfies the symplecticity and the reversibility 
(with $\lambda_r$ and $\lambda_r'$ interchanged):%
\footnote{
  $\omega(z,\pi)\equiv d\pi^I\wedge dz^I\bigr|_{T\calR}$
  is the induced symplectic form 
  for the embedding of the phase space $T\calR$ into $T\bbR^M$ 
  (see appendix \ref{sec:RATTLE_geometry}).
} 
\begin{align}
  &\bullet~\omega(z',\pi') = \omega(z,\pi), 
\\
  &\bullet~(z',\pi') = \Phi_{\Delta s}(z,\pi) 
  \Rightarrow (z, -\pi) = \Phi_{\Delta s}(z',-\pi'). 
\end{align}
The Hamiltonian is conserved to the order of $\Delta s^2$, 
i.e., $H(z',\pi')-H(z,\pi) = O(\Delta s^3)$. 

The HMC algorithm on $\calR$ then consists of the following three steps 
for a given initial configuration $z\in\calR$: 
\begin{description}

\item[Step 1.]
  Generate a vector $\tilde\pi=(\tilde{\pi}^I)\in T_z\bbR^M$ 
  from the Gaussian distribution 
  \begin{align}
    \frac{1}{(2\pi)^{M/2}} e^{- \tilde{\pi}^2/2 }, 
  \end{align}
  and project it onto $T_z\calR$ 
  to obtain an initial momentum $\pi=(\pi^I)\in T_z\calR$.

\item[Step 2.]
  Calculate $\Phi_{\Delta s}(z,\pi)$ 
  from \eqref{pihalf}--\eqref{constraint_piprime}. 
  We repeat this step $n$ times 
  to obtain $(z',\pi') = \Phi_{\Delta s}^n(z,\pi)$. 

\item[Step 3.]
  Update the configuration $z$ to $z'$ with a probability 
  \begin{align}
    \min\big( 1, e^{-H(z',\pi') + H(z,\pi)} \big). 
  \end{align}
\end{description}
The above process defines a stochastic process on $\calR$. 
One can show that 
its transition matrix $P(z'|z)$ satisfies the detailed balance condition 
(see appendix \ref{sec:RATTLE_geometry}): 
\begin{align}
  P(z'|z)\,e^{-V(z)} = P(z|z')\,e^{-V(z')},
\label{detailed_balance}
\end{align}

\section{HMC on the worldvolume}
\label{sec:wv_hmc}

In this section, 
we apply the general formalism in the previous section 
to the case where $\calR$ is the worldvolume of an integration surface. 
We first clarify the geometry of the worldvolume $\calR$ 
and then construct the HMC algorithm on $\calR$.

\subsection{Geometry of the worldvolume $\calR$}
\label{sec:wv_geometry}

\begin{figure}[ht]
  \centering
  \includegraphics[width=85mm]{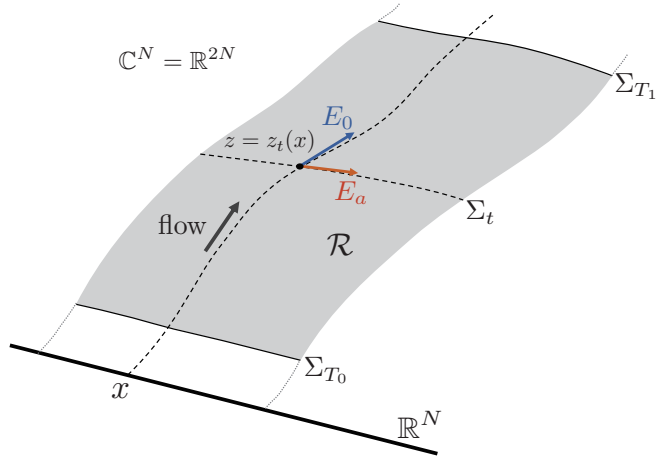} 
  \caption{
  \label{fig:TzR}
    A basis of $T_z \calR$, $\{E_\mu\} = \{E_0, E_a\}$. 
    $\calR$ is drawn one dimension less 
    than in Fig.~\ref{fig:calR}.
  }
\end{figure}

Recall that our worldvolume $\calR$ 
is an $N+1$-dimensional submanifold embedded in $\bbC^N=\bbR^{2N}$. 
As in the previous section, 
we again assume that $\calR \subset \bbR^{2N}$ is characterized 
by a set of independent constraint equations: 
$\phi^r(z)=0$ $(r=1,\cdots,N-1)$. 
We will often write a point $z=(z^i)\in\bbC^N$ $(i=1,\ldots,N)$ 
with real coordinates as%
\footnote{
  We use the same symbol for both complex and real coordinates 
  to avoid a mess of many symbols. 
  We will notify when one needs to specify which coordinates are implied. 
} 
\begin{align}
  z = (z^I) =
  \left( 
  \begin{array}{c}
    {\rm Re}\,z^i \\
    {\rm Im}\,z^i
  \end{array}
  \right)
  \in\bbR^{2N} \quad (I=1,\ldots,2N).
\end{align}
Since $\xi = (\xi^\mu) \equiv (t,x^a)$ specifies a point 
$z(\xi)=(z^I(\xi))\equiv (z_t^I(x))$ in $\calR$, 
we can use $\xi$ as coordinates of $\calR$. 
The flow equation \eqref{flow_zC} then takes the form 
\begin{align}
  \frac{\partial z^I}{\partial t} = \partial_{z^I} {\rm Re}\, S(z),\quad
  (z^I)|_{t=0} =   
  \left( 
  \begin{array}{c}
    x^i \\
    0
  \end{array}
  \right). 
\label{flow_zR}
\end{align}
Similarly, we write an $N$-dimensional complex vector 
$v=(v^i)\in \bbC^N$ $(i=1,\ldots,N)$ 
as a real vector%
\footnote{
  We define the inner product of two real vectors 
  $u=(u^I),\,v=(v^I) \in \bbR^{2N}$ by 
  $u\cdot v \equiv u^T v = u^I v^I$. 
  In terms of complex vectors $u=(u^i),\,v=(v^i)\in \bbC^N$, 
  the inner product is expressed as ${\rm Re}\,(u^\dagger v)$. 
  We do not distinguish the upper and lower indices for $I$. 
} 
\begin{align}
  v = (v^I) = \left(
  \begin{array}{c}
    {\rm Re}\,v^i\\
    {\rm Im}\,v^i\\
    \end{array}
  \right)
  \in\bbR^{2N} \quad (I=1,\ldots,2N).
\end{align}
The vectors $E_\mu=(E^I_\mu=\partial z^I/\partial \xi^\mu)$ 
form a basis of $T_z\calR$ 
(see Fig.~\ref{fig:TzR}), 
from which the induced metric $g_{\mu\nu}$ on $\calR$ 
is given by 
\begin{align}
  g_{\mu\nu} = E_\mu \cdot E_\nu. 
\end{align}
\noindent

\begin{figure}[ht]
  \centering
  \includegraphics[width=65mm]{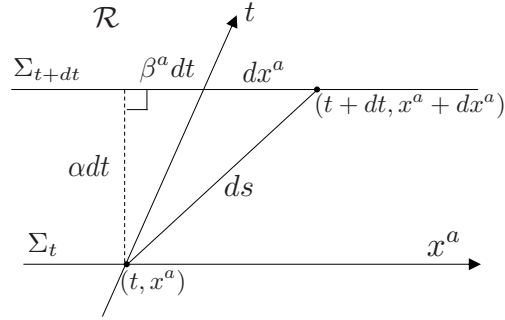} 
  \caption{
  \label{fig:ADM}
    Geometrical meaning of the ADM parametrization. 
    $\alpha\,dt$ is the geodesic distance 
    from a point $\xi=(t,x^a)$ on $\Sigma_t$ 
    to the surface $\Sigma_{t+dt}$. 
    $\beta^a dt$ shows that the time axis $t$ is tilted 
    with respect to the normal of $\Sigma_t$ by this amount in $x$-coordinates. 
    $\gamma_{ab}$ is the induced metric on $\Sigma_t$. 
    The geodesic distance $ds$ between two points $\xi=(t,x^a)$ 
    and $\xi+d\xi=(t+dt,x^a+dx^a)$ 
    is then obtained from Pythagorean theorem as in \eqref{ADM}. 
  }
\end{figure}

Since our worldvolume $\calR = \bigcup_t \Sigma_t$ 
is foliated by the antiholomorphic gradient flow, 
its intrinsic geometry should be best described (at least for physicists) 
by the Arnowitt-Deser-Misner (ADM) parametrization, 
for which the metric is expressed in the form 
\begin{align}
  ds^2 = g_{\mu\nu}(\xi)\,d\xi^\mu d\xi^\nu
  = \alpha^2(\xi)\,dt^2 
  + \gamma_{ab}(\xi)\,(dx^a + \shift^a(\xi) dt)(dx^b + \shift^b(\xi) dt).
\label{ADM}
\end{align}
Here, $\gamma_{ab}$ is the induced metric on $\Sigma_t$ 
(with its inverse matrix $\gamma^{ab}$), 
$\shift^a$ is the shift vector, 
and $\alpha$ is the lapse function,
\begin{align}
  \gamma_{ab} &\equiv E_a\cdot E_b,
\\
  \shift^a &\equiv \gamma^{ab}\,\shift_b \equiv \gamma^{ab}\,E_0\cdot E_b,
\\
  \alpha^2 &\equiv E_0\cdot E_0 - \gamma^{ab}\,\shift_a \shift_b
  =E_0\cdot (1-\gamma^{ab}\,E_a E_b^T)\,E_0. 
\end{align}
The inverse matrix of $(g_{\mu\nu})$ can be easily calculated to be 
\begin{align}
  (g^{\mu\nu}) = \left(
  \begin{array}{cc}
    1/\lapse^2& -\shift^b/\lapse^2 \\
    -\shift^a/\lapse^2 & \shift^a \shift^b/\lapse^2 + \gamma^{ab} \\
    \end {array}
  \right). 
\label{inverse_metric}
\end{align}
The geometrical meaning of the ADM parametrization is explained 
in Fig.~\ref{fig:ADM}. 
\noindent

 Since the flowed surfaces $\Sigma_t$ are determined 
by the flow equation, 
we can write down the explicit form of 
the basis $E_\mu=(E^I_\mu=\partial z^I/\partial \xi^\mu)$ 
of $T_z\calR$ as 
\begin{align}
  E^I_{0} = \partial_{z^I}{\rm Re}\,S(z), \quad
  E^I_{a} = \left(
  \begin{array}{c}
    {\rm Re}\,J^i_{~a} \\ {\rm Im}\,J^i_{~a}
  \end{array}\right),
\end{align}
where we have defined $J(\xi) = J(t,x) \equiv J_t(x)$. 
Thus, the induced metric $\gamma_{ab}$ can be directly expressed 
in terms of the Jacobian as%
\footnote{
\label{fn:Im_JdaggerJ}
  The second equality is a direct consequence of the identity 
  ${\rm Im}\, (J^\dagger J) = 0$, 
  which can be proved by a differential equation 
  $(d/dt)\,{\rm Im}\,(J^\dagger J) = 0$ 
  [as can be shown from \eqref{Jacobian}] 
  with the initial condition 
  ${\rm Im}\, (J^\dagger J)|_{t=0}  = 0$. 
} 
\begin{align}
  \gamma_{ab} = {\rm Re}\,(J^\dagger J)_{ab} = (J^\dagger J)_{ab}. 
\label{metric_g}
\end{align}
The lapse function $\alpha$ can be expressed 
as the length of the normal component of $E_0$: 
\begin{align}
  \lapse^2 = E_0^\perp\cdot E_0^\perp = (E_0^\perp)^2. 
\label{lapse_formula}
\end{align}
Here, the decomposition of $E_0$ to the tangential and normal components 
is given by 
\begin{align}
  &E_0 = E_0^\parallel + E_0^\perp,
\\
  &E_0^\parallel \equiv \Pi_{\Sigma_t}\, E_0, \quad 
  E_0^\perp \equiv (1-\Pi_{\Sigma_t})\, E_0, 
\end{align}
where $\Pi_{\Sigma_t}$ is the projector onto $T_z\Sigma_t$:%
\footnote{
  In actual calculations, 
  we do not need the explicit form of $\Pi_{\Sigma_t}$
  in projecting vectors in $T_z\bbR^{2N}$ onto $T_z\calR$ 
  (see subsection \ref{sec:wv_md}), 
  and thus do not have to calculate the Jacobian $J$. 
  A similar statement can be applied to the expressions below. 
} 
\begin{align}
  \Pi_{\Sigma_t} &\equiv \gamma^{ab} E_a E_b^T. 
\label{eq:proj_Sigma}
\end{align}
Note that the following relation holds: 
\begin{align}
  \shift^a E_a = E^\perp_0.
\label{shift_formula}
\end{align}
Then, by using \eqref{inverse_metric}, \eqref{lapse_formula}, 
\eqref{eq:proj_Sigma} and \eqref{shift_formula}, 
we see that the projector from $T_z\bbR^{2N}$ onto $T_z \calR$ 
is given by 
\begin{align}
  \Pi_{\calR} &\equiv g^{\mu\nu}E_\mu E_\nu ^T 
  = \frac{1}{(E_0^\perp)^2}E_0^\perp (E_0^\perp)^T + \Pi_{\Sigma_t}. 
\label{eq:proj_R}
\end{align}

\begin{figure}[ht]
  \centering
  \includegraphics[width=85mm]{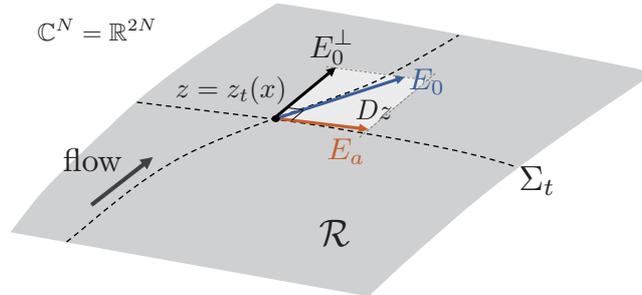} 
  \caption{
  \label{fig:volume}
    $E_0^\perp$ and the volume element.
  }
\end{figure}

In the ADM parametrization, 
the volume element of $\calR$ is given by 
(see Fig.~\ref{fig:volume})
\begin{align}
  \muR = \sqrt{g}\,d\xi = \alpha\,|\det J|\,dt\,dx.
\label{measure_R_2}
\end{align}
\noindent
Since the complex measure on $\Sigma_t$ is given 
by $dz_t=\det J\,dx$, 
we find that the reweighting factor \eqref{reweighting_factor} 
takes the form 
\begin{align}
  \denom(z) = \frac{dt\,dz_t}{\muR}\,e^{-i\,{\rm Im}\, S(z)} 
  = \alpha^{-1}(z)\,e^{i\varphi(z)-i\,{\rm Im}\, S(z)} \quad
  \Bigl(e^{i\varphi(z)} \equiv \frac{\det J}{|\det J|}\Bigr).
\label{phi_formula}
\end{align}
Note that the inverse lapse, $\alpha^{-1}(z)$, 
plays the role of the radius of $\denom(z)$. 

In appendix \ref{sec:gaussian}, 
in order to show the typical behaviors of 
various geometrical quantities near critical points at large flow times, 
we give explicit expressions of these quantities 
for the Gaussian case with the action 
\begin{align}
  S(x) = \frac{\beta}{2}\sum_{k=1}^N (x^k - i)^2. 
\end{align}
There, we find that 
$\alpha^{-1}(z)$ increases exponentially in flow time $t$ 
as $z^k=z^k(t,x)$ approaches the Lefschetz thimble 
at ${\rm Im}\,z^k=1$.

\subsection{Molecular dynamics on the worldvolume}
\label{sec:wv_md}

We first rewrite the Lagrange multiplier term 
$\lambda_r\, \partial \phi^r$ in \eqref{pidot} 
in a more convenient form. 
Note that $\lambda_r\, \partial \phi^r$ is normal to $\calR$,
and thus it satisfies 
\begin{align}
  (\lambda_r\,\partial \phi^r) \cdot E_0 = 0,
  \quad
  (\lambda_r\, \partial \phi^r) \cdot E_a = 0. 
\label{normal}
\end{align}
Since the vectors 
\begin{align}
  F_a \equiv \, 
  \left(\begin{array}{c}
    {\rm Im}{J^i}_a \\ -{\rm Re}\, {J^i}_a
  \end{array}\right)
  \quad (a=1,\ldots,N)
\end{align}
span the normal vector space $N_z\Sigma_t$ at $z\in\Sigma_t$,%
\footnote{
  $F_a$ form a basis of the normal space $N_z\Sigma_t$ 
  because $E_a \cdot F_b = - {\rm Im}\, (J^\dagger J)_{ab} = 0$ 
  (see footnote \ref{fn:Im_JdaggerJ}). 
  They can be written as $F_a = iE_a$  
  as complex vectors. 
} 
the second equation in \eqref{normal} means 
that $\lambda_r\, \partial \phi^r$ can be written as 
a linear combination of $F_a$ 
with new Lagrange multipliers $\lambda^a\in\bbR$ $(a=1,\ldots,N)$:%
\footnote{
  We here put the summation symbols 
  to stress the summation ranges for $r$ and $a$. 
} 
\begin{align}
  \sum_{r=1}^{N-1}\,\lambda_r\, \partial \phi^r = 
  \sum_{a=1}^N\,\lambda^a F_a. 
\end{align}
The first equation in \eqref{normal} is then treated as 
a constraint on $\lambda^a F_a$:%
\footnote{
  It is possible to solve the constraint \eqref{constraint_Fa} 
  as follows. 
  We first take a subset $\{F_r\}$ $(r=1,\ldots,N-1)$ of $\{F_a\}$, 
  whose elements are not parallel to $E_0$. 
  Then, we construct a basis of $N_z\calR$ by 
  $\tilde{F}_r\equiv F_r - (F_r\cdot E_0/(E_0)^2)E_0$, 
  and replace $\lambda^a F_a$ in \eqref{constraint_Fa} 
  by $\mu^r \tilde{F}_r$ 
  with new Lagrange multipliers $(\mu^r) \in \bbR^{N-1}$. 
} 
\begin{align}
  \lambda^a F_a\cdot E_0 = 0. 
\label{constraint_Fa}
\end{align}
 
From the above argument, 
we find that 
the RATTLE algorithm \eqref{pihalf}--\eqref{constraint_piprime} 
can be written as 
\begin{align}
  \pi_{1/2} &= \pi - \frac{\Delta s}{2}\, \partial V(z) - \lambda^a F_a(z),
\label{pihalf2}
\\
  z' &= z + \Delta s\, \pi_{1/2},
\label{zprime2}
\\
  \pi' &= \pi - \frac{\Delta s}{2}\, \partial V(z') - {\lambda'}^a F_a(z'),
\label{piprime2}
\end{align}
where $\lambda^a$ and ${\lambda'}^a$ are determined, respectively, 
so that the following constraints are satisfied: 
\begin{align}
  z' \in \calR \quad &\mbox{and} \quad 
  \lambda^a F_a(z) \cdot E_0(z) = 0,
\label{constraint_zprime2}
\\
  \pi' \in T_{z'} \calR \quad &\mbox{and} \quad 
  {\lambda'}^a F_a(z') \cdot E_0(z') = 0.
\label{constraint_piprime2}
\end{align}

The gradient of the potential, $\partial V(z)$, now takes the form 
\begin{align}
  \partial V(z) = \partial {\rm Re}\, S(z) + \wt'(t(z))\, \partial t(z).
\label{grad_V}
\end{align}
In order to define the gradient $\partial t(z)$ at $z\in\calR$, 
we regard $(\phi^r)$ as coordinates in the extra dimensions 
and construct $2N$ coordinates $(\xi^\mu, \phi^r)$ 
in the vicinity of $\calR$ in $\bbR^{2N}$. 
Then, one can show (see appendix \ref{sec:grad_t_proof})
that the gradient $\partial t(z)$ is given by 
\begin{align}
  \partial t(z) = \frac{1}{(E_0^\perp)^2}\, E_0^\perp 
  -\frac{E_r\cdot E_0^\perp}{(E_0^\perp)^2}\, \partial\phi^r(z)
  \quad
  (z\in\calR). 
\label{grad_t_general}
\end{align}
Since the last term is a linear combination of the gradients $\partial\phi^r(z)$ 
and thus can be absorbed into the Lagrange multiplier terms 
in \eqref{pihalf2} and \eqref{piprime2}, 
we can (and will) set $\partial t(z)$ to the form 
\begin{align}
  \partial t(z) = \frac{1}{(E_0^\perp)^2}\,E_0^\perp 
  \quad
  (z\in\calR). 
\label{grad_t}
\end{align}

\subsection{Solving the constraints in molecular dynamics}
\label{sec:wv_constraints}

In this subsection, 
we present numerical algorithms to solve the constraints 
\eqref{constraint_zprime2} and \eqref{constraint_piprime2}.

\noindent
\underline{\bf Solving \eqref{constraint_zprime2}}

\begin{figure}[ht]
\centering
\includegraphics[width=85mm]{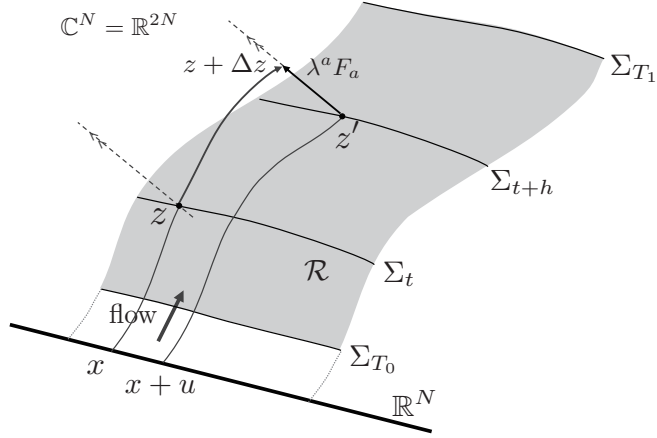} 
\caption{
  Finding $z'=z(\xi+\varepsilon)=z_{t+h}(x+u)$ on $\calR$ 
  from $z=z(\xi)=z_t(x)\in\calR$. 
\label{fig:RATTLE}
}
\end{figure}

The condition $z'\in\calR$ for a given $z=z(\xi)\in\calR$ 
is equivalent to the existence of an $N+1$-dimensional vector 
$\varepsilon=(\varepsilon^\mu)=(h,u^a)$ 
such that $z'=z(\xi+\varepsilon)=z_{t+h}(x+u)$. 
Thus, \eqref{constraint_zprime2} can be solved 
by finding a solution to $2N+1$ equations 
\begin{align}
  f^P(w) = 0 \quad (P=0,1,\ldots,2N)
\label{f=0}
\end{align}
for $2N+1$ unknowns 
$w=(w^P)=(\varepsilon^\mu, \lambda^a) \in \bbR^{2N+1}$ 
(see Fig.~\ref{fig:RATTLE}),  
where 
\begin{align}
  f^0(w) &\equiv \lambda^a\, F_a(\xi)\cdot E_0(\xi) 
  = - \lambda^a\, {\rm Im} \bigl[ \partial_{z^i} S(z(\xi))\,
  {J^i}_a(\xi)\bigr], 
\label{f0}
\\
  f^I(w) &\equiv z^I(\xi+\varepsilon) - z^I(\xi) 
  - \Delta z^I + \lambda^a F_a^I(\xi) \quad
  (I=1,\ldots,2N)
\label{fI}
\end{align}
with 
\begin{align}
  \Delta z \equiv \Delta s\,\pi 
  - \frac{\Delta s^2}{2}\, \partial V(z) . 
\end{align}
\noindent
Adopting Newton's method to find a solution, 
we iteratively update the vector $w=(w^P)=(\varepsilon^\mu,\lambda^a)$ 
as $w \to w + \Delta w$, 
where $\Delta w$ is a solution of the linear equation 
\begin{align}
  \frac{\partial f^P(w)}{\partial w^Q} \Delta w^Q = - f^P(w). 
\label{newton}
\end{align}
The matrix elements $\partial f^P/\partial w^Q$ are easily found to be 
\begin{align}
  \Big(\frac{\partial f^P(w)}{\partial w^Q}\Big)
  &= \left[ 
    \begin{array}{ccc}
      0 & 0 & F_a(\xi) \cdot E_0(\xi) \\
      E^I_0(\xi+\varepsilon) & E^I_a(\xi+\varepsilon) & F^I_a(\xi) \\
    \end{array}
  \right] \nonumber\\ 
  &= \left[ 
    \begin{array}{ccc}
      0 & 0 &  -{\rm Im}\,\bigl[\partial_{z^i}S(z(\xi))\,{J^i}_a(\xi)\bigr]
    \\
      {\rm Re}\, \bigl[\partial_{z^i} S(z(\xi+\varepsilon))\bigr]
        & {\rm Re}\,{J^i}_a(\xi+\varepsilon)
            & -{\rm Im}\,{J^i}_a(\xi) 
    \\
      -{\rm Im}\, \bigl[\partial_{z^i} S(z(\xi+\varepsilon))\bigr]
        & {\rm Im}\,{J^i}_a(\xi+\varepsilon)
            & {\rm Re}\,{J^i}_a(\xi) \\
    \end{array}
  \right]. 
\label{dfdw}
\end{align}

When the DOF $(=N)$ is small, 
\eqref{newton} can be solved by a direct method 
such as the LU decomposition of \eqref{dfdw}, 
for which we integrate the second flow equation \eqref{Jacobian} 
to know the value of $J$. 
When the DOF is large, 
the computation of $J$ becomes expensive, 
and we can instead use an iterative method 
such as GMRES \cite{Saad:1983} or BiCGStab \cite{Vorst:1990}, 
for which we do not need to compute the matrix elements of $J$ 
as in \cite{Alexandru:2017lqr}. 
To see this, 
we first note that 
the right hand side of \eqref{newton} can be written 
in terms of complex vectors as 
\begin{align}
  f^0(w) &= - {\rm Im}\, \bigl[\partial_{z^i} S(z(\xi))\,
  {J^i}_a(\xi)\lambda^a \bigr], 
\label{f'}
\\
  f^i(w) &= z^i(\xi+\varepsilon) - z^i(\xi) - \Delta z^i 
  + i {J^i}_a(\xi)\lambda^a. 
\label{fi'}
\end{align}
The left hand side of \eqref{newton} can also be written as 
\begin{align}
  \frac{\partial f^0(w)}{\partial w^N}\, \Delta w^N 
  &= - {\rm Im}\,\bigl[\partial_{z^i} S(z(\xi))\,
  {J^i}_a(\xi) \Delta\lambda^a \bigr],
\label{df0} 
\\
  \frac{\partial f^i(w)}{\partial w^N}\, \Delta w^N 
  &= [\partial_{z^i} S(z(\xi+\varepsilon))]^\ast 
    + {J^i}_a(\xi+\varepsilon) \Delta u^a
    + i{J^i}_a(\xi) \Delta \lambda^a. 
\label{dfi}
\end{align}
We thus see that in the above equations, 
$J$ appears only in the form 
${J^i}_a(\xi) v^a$ or ${J^i}_a(\xi+\varepsilon) v^a$
with a real vector $v=(v^a)\in\bbR^N$. 
The former can be evaluated 
from the solution to the flow equation \eqref{flow_zC} 
and the following equation [see \eqref{Jacobian}]:
\begin{align}
  \dot{v}_t^i = [H_{ij}(z_t)\,v_t^j]^\ast,\quad v^i_{t=0}=v^i,
\label{vector_flow}
\end{align}
by setting ${J^i}_a(\xi) v^a=v_t^i$. 
The latter is obtained in a similar way, 
by replacing $t\to t+h$ and $x^a\to x^a+u^a$. 
We thus find that the both hand sides of \eqref{newton} 
can be calculated without computing the matrix elements of $J$. 

\noindent
\underline{\bf Solving \eqref{constraint_piprime2}}

We first note that 
solving the constraint \eqref{constraint_piprime2} 
is equivalent to projecting the vector  
\begin{align}
  \tilde{\pi}' &\equiv \pi - \frac{\Delta s}{2}\, \partial V(z') 
\label{tilde_pi}
\end{align}
onto $T_{z'}\calR$ [see \eqref{eq:proj_R}]: 
\begin{align}
  \pi' &\equiv \Pi_{\calR}(z')\, \tilde{\pi}' 
  = \frac{E_0^\perp(z')\cdot \tilde{\pi}'}{(E_0^\perp(z'))^2}\,E_0^\perp(z')
  + \Pi_{\Sigma_t}(z')\, \tilde{\pi}'. 
\label{projection_pi'}
\end{align}
Here, $E_0^\perp(z')$ can be computed as a complex vector to be 
\begin{align}
  E_0^\perp(z') = i J(z')\, {\rm Im}\, 
  \bigl( J^{-1}(z')\, [\partial S(z')]^\ast \bigr). 
\label{E0perp_C}
\end{align}
The second term in \eqref{projection_pi'} 
can also be computed as
\begin{align}
  \Pi_{\Sigma_t}(z')\, \tilde{\pi}' 
  = J(z')\, {\rm Re}\, \bigl( J^{-1}(z')\, \tilde{\pi}' \bigr). 
\label{Ppi_C}
\end{align}

The expressions \eqref{E0perp_C} and \eqref{Ppi_C} 
can again be evaluated either by a direct method 
with the computation of $J(z')$, 
or by an iterative method 
without computing $J(z')$ 
as in \cite{Alexandru:2017lqr}. 
When the iterative method is used, 
the inversion $J^{-1}(z')\, c$ 
is obtained 
for a given complex vector $c=(c^i)\in\bbC^N$ 
by looking for vectors $a=(a^a),\,b=(b^a)\in\bbR^N$ iteratively 
such that 
\begin{align}
  c = J(z')\,a + i J(z')\,b, 
\label{inversion_iterative}
\end{align}
where $J(z')\,a$ and $J(z')\,b$ are evaluated 
by integrating the flow equation \eqref{vector_flow} 
with the initial condition $v_{t=0}=a$ and $v_{t=0}=b$, respectively. 
The multiplication $J(z')\, v$ 
are again calculated by integrating \eqref{vector_flow}. 
Therefore, the projection \eqref{projection_pi'}
can be performed without computing $J(z')$. 

Note that every time we evaluate $\partial V(z)$ 
[eq.~\eqref{grad_V}], 
we need $\partial t(z)$ and thus $E_0^\perp(z)$ [see \eqref{grad_t}]. 
$E_0^\perp(z)$ can be calculated from \eqref{E0perp_C} 
by replacing $z'$ with $z$.

\subsection{Construction of $\wt(t)$}
\label{sec:weight}

In this subsection, 
we present a prescription to construct a weight function $e^{-\wt(t)}$ 
in \eqref{integral_R_intro} [or in \eqref{potential}]
so that it gives an almost uniform distribution with respect to $t$. 
The key is that, for a given weight $e^{-\wt(t)}$, 
the probability to find a configuration at $t$ 
is proportional to 
\begin{align}
  Z(t; \wt) \equiv e^{-\wt(t)} \int_{\bbR^N} dx \,
  \lapse(t,x) |\det J(t,x)| e^{-{\rm Re}\,S(z(t,x))}. 
\label{ideal_weight}
\end{align}
Thus, when a weight $e^{-\wt(t)}$ 
does not give a uniform distribution of $t$, 
the desired weight can be obtained by (see e.g., \cite{Wang:2000fzi}) 
\begin{align}
  \wt^{\rm (new)}(t) = \wt(t) + \ln Z(t;\wt) + \mbox{const}, 
\end{align}
because then $Z(t;\wt^{\rm (new)})$ becomes constant in $t$: 
\begin{align}
  Z(t;\wt^{\rm (new)}) &= 
  e^{-\wt^{\rm (new)}(t)} \int_{\bbR^N} dx \,
  \lapse(t,x)\, |\det J(t,x)|\, e^{-{\rm Re}\,S(z(t,x))}
\nonumber
\\
  &= {\rm const}\,\frac{e^{-\wt(t)}}{Z(t;\wt)} \cdot 
  e^{\wt(t)}\,Z(t;\wt) = \mbox{const}.
\end{align}

\begin{figure}[ht]
  \centering
  \includegraphics[width=85mm]{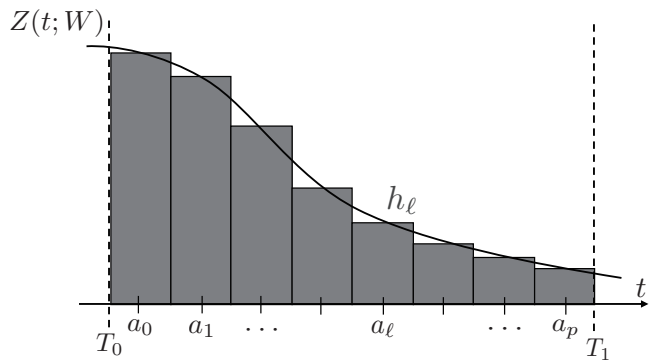} 
  \caption{
  \label{fig:Z_histo}
    Estimating $Z(t;\wt)$ from $\{h_\ell\}$. 
  }
\end{figure}

Of course, 
the above procedure is possible 
only when we know the values of $Z(t;\wt)$ explicitly, 
which is usually not the case. 
However, we can estimate $Z(t;\wt)$ 
from the histogram of flow times $\{t\}$. 
To be more specific, 
we first divide the interval $[T_0,T_1]$ into $p+1$ bins, 
$I_\ell \equiv [\ell h, (\ell+1)h]$ $(\ell=0,\ldots,p)$ 
with $h \equiv (T_1 - T_0)/(p+1)$, 
and generate a certain number ($\equiv n_{\rm tune}$) of configurations 
by using $V(z) = {\rm Re}\, S(z) + \wt(t(z))$ as the potential. 
The numbers $h_\ell$ of configurations inside the bins $I_\ell$ 
give a rough estimate of the functional form of $Z(t; \wt)$ 
up to a normalization factor (see Fig.~\ref{fig:Z_histo}). 
\noindent 
Then, from the histogram $h_\ell$ $(\ell=0,\ldots,p)$, 
we calculate
\begin{align}
  \wt^{\rm (new)}_\ell = \wt(a_\ell) + \ln h_\ell,
\end{align}
and construct a function $W^{\rm (new)}(t)$ 
to be approximated by a polynomial satisfying 
$W^{\rm (new)}(a_\ell)=W^{\rm (new)}_\ell$ 
with $a_\ell\equiv (\ell+1/2)h$. 

In general,
the minimum-order polynomial that have values $b_\ell$ 
at $a_\ell= (\ell+1/2)h$ 
is given by the Lagrange interpolation of the form
\begin{align}
  L(t;\{b_\ell\}) \equiv
  \sum_{\ell=0}^p 
  \frac{\prod_{m\neq\ell} (t-a_m)}
    {\prod_{m\neq\ell} (a_\ell-a_m)}\,b_\ell 
  = \sum_{\ell=0}^p \Bigl[ \prod_{m<\ell} (t-a_\ell) \Bigr]\,
  \frac{\Delta^\ell b_0}{\ell! \, h^\ell }, 
\label{lagrange_interpolation}
\end{align}
where, for an array $\{v_0,v_1,v_2,\ldots\}$, 
we define $\Delta v_\ell\equiv v_{\ell+1}-v_\ell$, 
so that
\begin{align}
  \Delta^\ell v_0 = \sum_{k=0}^\ell (-1)^k
  \binom{\ell}{k}\, v_{\ell-k}. 
\end{align}
Using this polynomial, 
we define%
\footnote{
  In the calculation below, 
  we put two additional terms 
  in \eqref{lagrange_interpolation} 
  to prevent the function from changing drastically near boundaries. 
  The above $L(t;\{b_\ell\})$ is then replaced by 
  \begin{align}
    L(t;\{b_\ell\}) 
    &\equiv ({\rm const.}t + {\rm const.}) \prod_{\ell=0}^p (t-a_\ell)
    +\sum_{\ell=0}^p \Bigl[ \prod_{m<\ell} (t-a_\ell) \Bigr]\,
    \frac{\Delta^\ell b_0}{\ell! \, h^\ell }. 
  \nonumber
  \end{align}
  The constants are determined by the conditions 
  $L'(T_0;\{\wt_\ell\}) = c_0$ and 
  $L'(T_1;\{\wt_\ell\}) = c_1$. 
  We set $c_0 = 1.2 \times \min_\ell(0, \Delta \wt_\ell / h)$ 
  and $c_1 = 0.01$ 
  in the calculation below. 
} 
\begin{align}
  \wt^{\rm (new)}(t) \equiv L(t;\{\wt^{\rm (new)}_\ell\}). 
\label{W^new}
\end{align}

Since the estimate of $Z(t;\wt)$ from the histogram $\{h_\ell\}$ 
includes statistical errors, 
we use an iterative algorithm 
to update $\{\wt_\ell\}$ 
until an almost uniform distribution is obtained: 
\begin{itemize}

\item
  Initialize $\{\wt_\ell\}$ with appropriate values 
  (e.g., $\wt_\ell^{(0)} = 0$). 

\item
  From an array $\{\wt^{(k)}_\ell\}$, 
  construct an order-$(p+1)$ polynomial $L(t;\{W^{(k)}_\ell\})$, 
  and set $W^{(k)}(t)\equiv L(t;\{W^{(k)}_\ell\})$. 

\item
  Generate $n_{\rm tune}$ configurations 
  with the potential $V(z)={\rm Re}\,S(z)+W^{(k)}(t(z))$, 
  and record the numbers $h_\ell^{(k)}$ of configurations  
  in the intervals $I_\ell$. 

\item
  Update $\{\wt_\ell\}$ 
  as $\wt_\ell^{(k+1)} \equiv \wt_\ell^{(k)} 
  + \ln (h_\ell^{(k)} + \epsilon_c)$. 
  Here, $\epsilon_c$ is a cutoff 
  to avoid the divergence arising when $h^{(k)}_\ell=0$. 

\item
  Terminate the iteration when the histogram becomes almost flat. 
  We use the following stopping condition: 
  \begin{align}
    \frac{1}{p}\sum_{l=0}^{p-1} 
    \Big[ \frac{h_{l+1}-h_l}{(h_{l+1}+h_l)/2} \Big]^2 < \delta^2. 
  \label{weight_convergence}
  \end{align}
\end{itemize}
In the calculation below, 
we set $p=7$, $n_{\rm tune}=(p+1)\times 200 = 1,600$, 
$\epsilon_c=0.01$, and $\delta^2 = 0.2$.

\subsection{Summary of the HMC algorithm on the worldvolume}
\label{sec:summary}

We summarize the HMC algorithm 
for a given initial configuration $z \in \calR$. 
\begin{description}

\item[Step 1.]
  Generate $\tilde\pi=(\tilde{\pi}^I)$ 
  from the Gaussian distribution, 
  and project it onto $T_z\calR$ 
  to obtain an initial momentum $\pi=(\pi^I)$: 
  $\pi = \Pi_\calR(z)\,\tilde\pi$. 

\item[Step 2.]
  Calculate $(z,\pi)\to\Phi_{\Delta s}(z,\pi)$ 
  with \eqref{pihalf2}--\eqref{constraint_piprime2}. 
  The gradient $\partial V(z)$ takes the form \eqref{grad_V}, 
  where $\partial t(z)$ is given by \eqref{grad_t} 
  and $W(t)$ is determined from test runs 
  by using the iterative algorithm given in subsection \ref{sec:weight}.
  The first constraint \eqref{constraint_zprime2} is solved 
  by finding a root of the functions \eqref{f0} and \eqref{fI}, 
  and the second constraint \eqref{constraint_piprime2} is solved 
  by calculating \eqref{tilde_pi} and \eqref{projection_pi'}. 
  We repeat this process $n$ times 
  to obtain $(z',\pi') = \Phi^n_{\Delta s}(z,\pi)$. 

\item[Step 3.]
  Update the configuration $z$ to $z'$ with a probability 
  \begin{align}
    \min\big( 1, e^{-H(z',\pi') + H(z,\pi)} \big). 
  \end{align}
\end{description}
Upon measurement, 
we further compute the reweighting factor $\denom(z)$ 
[see \eqref{phi_formula}], 
which requires the phase 
$e^{i\varphi(z)}=\det J(z)/|\det J(z)|$, 
that is evaluated 
by first solving \eqref{Jacobian} to get $J(z)$ 
and then computing its determinant. 
The lapse function $\lapse(z) = |E_0^\perp(z)|$ 
is already obtained in the preceding molecular dynamics step 
(Step 2). 

\begin{figure}[ht]
  \centering
  \includegraphics[width=85mm]{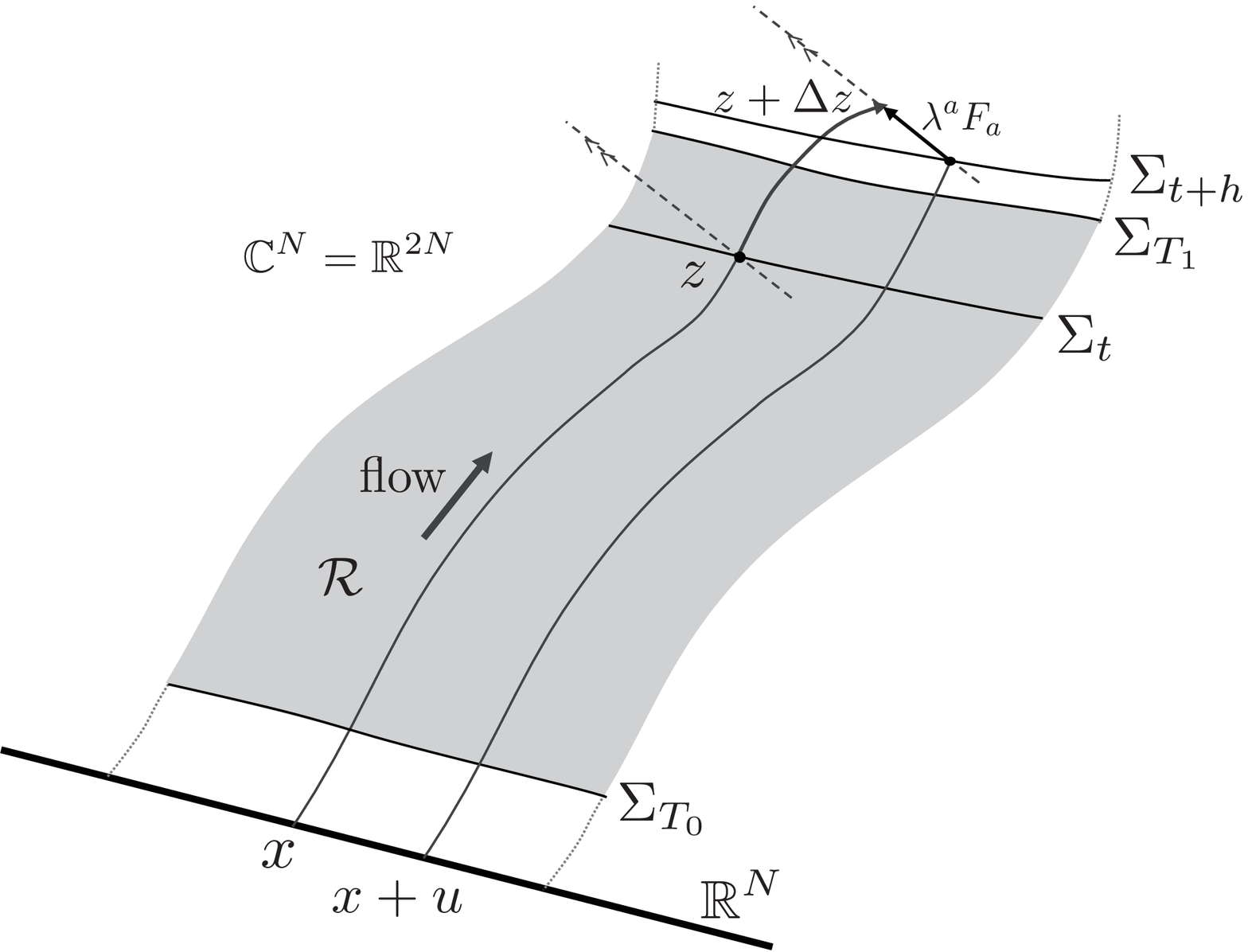} 
  \caption{
  \label{fig:boundary}
    A transition passing over the boundary at $T_1$. 
  }
\end{figure}

In the course of molecular dynamics (Step 2),  
it sometimes happens that 
the equation $(z',\pi') = \Phi_{\Delta s}(z,\pi)$ 
does not have a solution 
because $z+\Delta z$ passes over the boundary of $\calR$ 
(see Fig.~\ref{fig:boundary}). 
\noindent 
Here, the boundary in the $t$ direction is given by $T_0, \,T_1$, 
while that in the $x$ direction by zeros of $e^{-S(z)}$. 
When a solution is not found near a boundary, 
we replace the operator $\Phi_{\Delta s}$ 
by the {\it momentum reflection} $\mathcal{T}$: 
\begin{align}
  \mathcal{T}(z,\pi) \equiv (z,\pi'),
\end{align}
where for $\pi$ which is expanded with $E^\mu\equiv g^{\mu\nu} E_\nu$ 
as $\pi=\eta_0 E^0+\eta_a E^a$, 
the reflected momentum $\pi'$ is defined 
by $\pi'\equiv -\eta_0 E^0+\eta_a E^a=\pi - 2\eta_0 E^0$, 
i.e.,
\begin{align}
  \pi' \equiv \pi - 2\,\frac{E_0\cdot\pi}{{E_0^\perp}^2}\,E_0^\perp. 
\label{momentum_reflection}
\end{align}
This preserves the reversibility 
and the phase-space volume element 
because the induced symplectic form is given by 
$\omega=d\eta_\mu\wedge d\xi^\mu=d\eta_0\wedge d\xi^0+d\eta_a\wedge d\xi^a$ 
(see appendix \ref{sec:RATTLE_geometry}). 
However, this can change the value of the Hamiltonian. 
The change comes only from the difference 
between the norms of momenta $\pi$ and $\pi'$, 
and its effect is absorbed in the probability at the Metropolis test 
in Step 3 above, 
so that the detailed balance condition \eqref{detailed_balance} 
still holds. 
If the change is larger than a prescribed value 
(e.g., if $e^{-|\Delta H|}=e^{-|{\pi'}^2-\pi^2|/2}<0.8$), 
we instead use the {\it momentum flip} $\Psi$ \cite{Fukuma:2019uot}: 
\begin{align}
  \Psi(z,\pi) \equiv (z,-\pi). 
\label{momentum_flip}
\end{align}
Since the replacement of $\Phi_{\Delta s}$ by $\mathcal{T}$ or $\Psi$ 
preserves the phase-space volume element and the reversibility, 
the detailed balance condition \eqref{detailed_balance} still holds.%
\footnote{
  In practice, 
  we check the reversibility at every step of molecular dynamics, 
  $(z,\pi)\to(z',\pi')=\Phi_{\Delta s}(z,\pi)$, 
  by monitoring that the time-reversed process 
  $(\tilde z,-\tilde\pi)\equiv\Phi_{\Delta s}(z',-\pi')\,
  (=\Phi_{\Delta s}\circ\Psi\circ\Phi_{\Delta s}(z,\pi))$
  correctly gives $(z,-\pi)$. 
  In the calculation below, 
  we require that $|\tilde z-z|/\sqrt{N}<10^{-5}$. 
  If this condition is not met, 
  we replace $\Phi_{\Delta s}$ by $\mathcal{T}$ or $\Psi$ 
  as in the case where a transition passes over a boundary. 
} 

\subsection{Estimation of observables}
\label{sec:estimation}

We first recall that 
the boundary flow times ($T_0$ and $T_1$) 
can be chosen arbitrarily due to Cauchy's theorem. 
In practice, 
$T_0$ must be set sufficiently small 
in order to keep in $\calR$ 
a region that is free from multimodality 
(to be set to $T_0=0$ when the multimodal problem is absent there). 
On the other hand, $T_1$ must be taken sufficiently large 
in order to keep a region where the sign problem is resolved, 
but, at the same time, 
$T_1$ should not be set too large 
in order to avoid introducing an unnecessarily large computational time. 

When estimating observables, 
we take a subinterval in $[T_0,T_1]$  
(to be denoted by $[\hat{T}_0,\hat{T}_1]$). 
Namely, for a sample of configurations 
that are generated in the range $[T_0,T_1]$ 
(with sample size $N_{\rm conf}$), 
we construct a subsample $\{z^{(k)}\}$ $(k=1,\dots,\hat{N}_{\rm conf})$ 
taking configurations from the interval $[\hat{T}_0,\hat{T}_1]$ 
$(T_0 \leq \hat{T}_0 < \hat{T}_1 \leq T_1)$, 
and take a ratio of sample averages over this subsample,%
\footnote{
  $\hat{N}_{\rm conf}=N_{\rm conf}(\hat{T}_0,\hat{T}_1)$ 
  is the number of configurations in $[\hat{T}_0,\hat{T}_1]$. 
  The total number of configurations corresponds to 
  $N_{\rm conf} = N_{\rm conf}(T_0,T_1)$.
} 
\begin{align}
  \bar{\mathcal{O}}(\hat{T}_0,\hat{T}_1) \equiv
  \frac{(1/\hat{N}_{\rm conf})\,\sum_k\denom(z^{(k)}) \,
  \mathcal{O}(z^{(k)})}
  {(1/\hat{N}_{\rm conf})\,\sum_k\denom(z^{(k)})}.
\label{estimation_formula}
\end{align}
$\hat T_0$ now should be set sufficiently large 
in order to exclude a region that is contaminated by the sign problem. 
Note that $\hat T_0$ and $\hat T_1$ should be set enough apart 
in order to maintain a sufficient size for the subsample. 
Then, 
if the original range $[T_0,T_1]$ is properly chosen as above, 
and if the system is well close to global equilibrium, 
there must be a region in two-parameter space $(\hat{T}_0,\hat{T}_1)$ 
such that the estimations $\bar{\mathcal{O}}(\hat{T}_0,\hat{T}_1)$ 
are stable against the variation of the estimation ranges 
(i.e., estimates change only within statistical errors).

The whole process of the WV-TLTM thus proceeds as follows: 
\begin{enumerate}

\item
  Choose a sufficiently small $T_0$ and a sufficiently large $T_1$ 
  to tame both sign and multimodal problems. 

\item
  Construct a weight function $e^{-\wt(t)}$ 
  such that the distribution of $t$ becomes almost uniform 
  (see subsection \ref{sec:weight} for more details). 

\item
  Use the HMC algorithm in subsection \ref{sec:summary} 
  to generate configurations 
  in the range $[T_0,T_1]$
  from the distribution 
  $\propto e^{-V(z)}$ 
  with $V(z) = {\rm Re}\,S(z) + \wt(t(z))$. 

\item
  For the obtained full sample, 
  vary the estimation range $[\hat{T}_0,\hat{T}_1]$, 
  looking for a stable region (plateau) 
  in the two-parameter space $(\hat{T}_0,\hat{T}_1)$ 
  that give the same estimate 
  $\bar{\mathcal{O}}(\hat{T}_0,\hat{T}_1)$ 
  within statistical errors. 

\item
  Choose a point $(\hat{T}_0,\hat{T}_1)$ from the plateau 
  and take the corresponding $\bar{\mathcal{O}}(\hat{T}_0,\hat{T}_1)$ 
  as the estimate of $\langle\mathcal{O}(x)\rangle$. 
  The error of estimation is read from the statistical error 
  for the chosen subsample.
  
\end{enumerate}
%

\section{Application to a chiral random matrix model}
\label{sec:application}

In this section, 
to confirm that the WV-TLTM works correctly, 
we apply the WV-TLTM to a chiral random matrix model, 
the Stephanov model \cite{Stephanov:1996ki,Halasz:1998qr}. 
We show that the algorithm correctly reproduces exact results, 
solving both sign and multimodal problems.

\subsection{The model}
\label{sec:model}

The Stephanov model is a large $N$ matrix model 
that approximates QCD at finite density. 
For $N_f$ quarks with equal mass $m$, 
the partition function is given by the following integral 
over $n\times n$ complex matrices $X=(X_{ij}=x_{ij}+i\,y_{ij})$:
\begin{align}
  Z_n^{N_f} = e^{n\mu^2}\int d^2 X\,e^{-S(X,X^\dagger)}
  \equiv e^{n\mu^2}\int d^2 X\,e^{-n\,{\rm tr}\,X^\dagger X}\,
  {\det}^{N_f}(D+m).
\label{stephanov_partition}
\end{align}
Here, $D+m$ represents the Dirac operator 
in the chiral representation 
and takes the form
\begin{align}
  D + m \equiv \left(\begin{array}{cc}
    m\,1_n & i\,(X+C)\\
    i\,(X^\dagger+C) & m\,1_n
  \end{array}\right),
\end{align}
where 
\begin{align}
  i\,C \equiv \left(\begin{array}{cc}
    (\mu+i\tau)\,1_{n/2} & 0\\
    0 & (\mu-i\tau)\,1_{n/2}
  \end{array}\right).
\end{align}
The parameters $\mu$ and $\tau$ correspond to the chemical potential 
and the temperature, respectively \cite{Stephanov:1996ki,Halasz:1998qr}.
The number of DOF is $N=2n^2$, 
which may be compared with the DOF of the $SU(N_c)$ gauge field 
on the lattice of linear size $L$ 
as $N=4(N_c^2-1) L^4 $. 

For the case $N_f=1$, 
the partition function at finite $n$ can be written 
as an integral over a single variable: 
\begin{align}
  Z_n^{N_f=1}
  = n\,e^{n(\mu^2-m^2)}\,
  \int_0^\infty d\rho\,e^{-n\rho}\,I_0(2nm\sqrt{\rho})\,
  \bigr[(\rho-\mu^2+\tau^2)^2+(2\mu\tau)^2\bigr]^{n/2},
\end{align}
where $I_k(x)$ $(k=0,1,2,\cdots)$ 
are the modified Bessel functions of the first kind.
Then, the chiral condensate is expressed as 
\begin{align}
  \langle \bar\psi \psi \rangle
  &\equiv \frac{1}{2n}\,\frac{\partial}{\partial m}\,
  \ln Z_n^{N_f=1}
\nonumber
\\
  &= -m + 
  \frac{
    \int_0^\infty d\rho\,e^{-n\rho}\,I_1(2nm\sqrt{\rho})\,\sqrt{\rho}\,
    \bigr[(\rho-\mu^2+\tau^2)^2+(2\mu\tau)^2\bigr]^{n/2}
  }{
    \int_0^\infty d\rho\,e^{-n\rho}\,I_0(2nm\sqrt{\rho})\,
    \bigr[(\rho-\mu^2+\tau^2)^2+(2\mu\tau)^2\bigr]^{n/2}
  }.
\end{align}
Similarly, the number density is expressed as 
\begin{align}
  \langle \psi^\dagger \psi \rangle
  &\equiv \frac{1}{2n}\,\frac{\partial}{\partial \mu}\,
  \ln Z_n^{N_f=1}
\nonumber
\\
  &= \mu - \mu\,
  \frac{
    \int_0^\infty d\rho\,e^{-n\rho}\,I_0(2nm\sqrt{\rho})\,
    \bigr[(\rho-\mu^2+\tau^2)^2+(2\mu\tau)^2\bigr]^{n/2-1}\,
    (\rho-\mu^2-\tau^2)
  }{
    \int_0^\infty d\rho\,e^{-n\rho}\,I_0(2nm\sqrt{\rho})\,
    \bigr[(\rho-\mu^2+\tau^2)^2+(2\mu\tau)^2\bigr]^{n/2}
  }.
\end{align}

We apply the WV-TLTM to this model, 
by complexifying the real and imaginary parts 
($x_{ij}$ and $y_{ij}$) separately, 
and by considering the antiholomorphic gradient flow 
with respect to the action given in \eqref{stephanov_partition}. 
We estimate the chiral condensate and the number density 
using the formulas  
\begin{align}
  \langle \bar\psi \psi \rangle
  &=\frac{1}{2n}\,\langle {\rm tr}\,(D+m)^{-1} \rangle,\quad
\label{stephanov_vev1}
\\
  \langle \psi^\dagger \psi \rangle
  &=\mu+\frac{1}{2n}\,\Bigl\langle 
  {\rm tr}\,\Bigl[(D+m)^{-1}
  \left(\begin{array}{cc}
    0 & 1_n \\
    1_n & 0
  \end{array}\right)\Bigr]
  \Bigr\rangle.
\label{stephanov_vev2}
\end{align}
It is convenient to introduce the matrices 
\begin{align}
  A \equiv X+C,\quad B \equiv X^\dagger+C,\quad
  K \equiv (BA+m^2)^{-1}, 
\end{align}
with which $D+m$ and $(D+m)^{-1}$ are expressed as%
\footnote{
  Note that 
  $(AB+m^2)^{-1} = A K A^{-1}
  = B^{-1} K B = (1/m^2)\,(1-AKB)$.
} 
\begin{align}
  D+m &= \left(\begin{array}{cc}
    m\,1_n & i\,A \\
    i\,B & m\,1_n
  \end{array}\right),
\\
  (D+m)^{-1} &=  \left(\begin{array}{cc}
    m\,(AB+m^2)^{-1} & -i\,A (BA+m^2)^{-1} \\
    -i\,(BA+m^2)^{-1} B & m\,(BA+m^2)^{-1}
  \end{array}\right)
\nonumber
\\ 
  &=  \left(\begin{array}{cc}
    m\,A K A^{-1} & -i\,A K \\
    -i\,K B & m\,K
  \end{array}\right).
\end{align}
The flow equation is then written only with $A,\,B,\,K$, 
and the expectation values \eqref{stephanov_vev1} and \eqref{stephanov_vev2} 
are estimated from the expressions
\begin{align}
  \langle \bar\psi \psi \rangle
  &=\frac{m}{n}\,\langle {\rm tr\,}K \rangle,\quad
\\
  \langle \psi^\dagger \psi \rangle
  &=\mu - \frac{i}{2n}\,\langle {\rm tr\,}K (A+B) \rangle.
\end{align}

\subsection{Setup in the simulation}
\label{sec:setup}

We summarize the setup in the simulation. 
We set $n=10$, $m=0.004$, $\tau=0$, 
and estimate 
the chiral condensate $\langle\bar\psi\psi\rangle$ 
and the number density $\langle\psi^\dagger\psi\rangle$
as functions of $\mu=0.4,\ldots,0.8$.%
\footnote{
  Since the lattice size is small, 
  we adopt the direct method in the HMC algorithm; 
  we compute $J$ by integrating the flow equation \eqref{Jacobian} 
  and use the LU decomposition in the inversion processes. 
  The computation of $J$ is not necessary 
  if we adopt the iterative method 
  (see subsection \ref{sec:wv_constraints}). 
} 

\begin{figure}[ht]
  \centering
  \includegraphics[width=70mm]{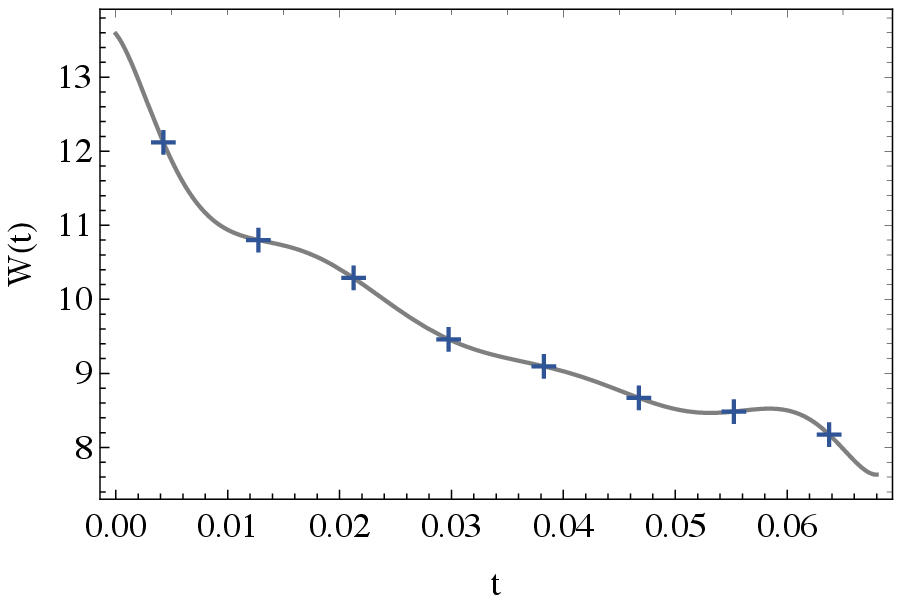} \hspace{5mm}
  \includegraphics[width=70mm]{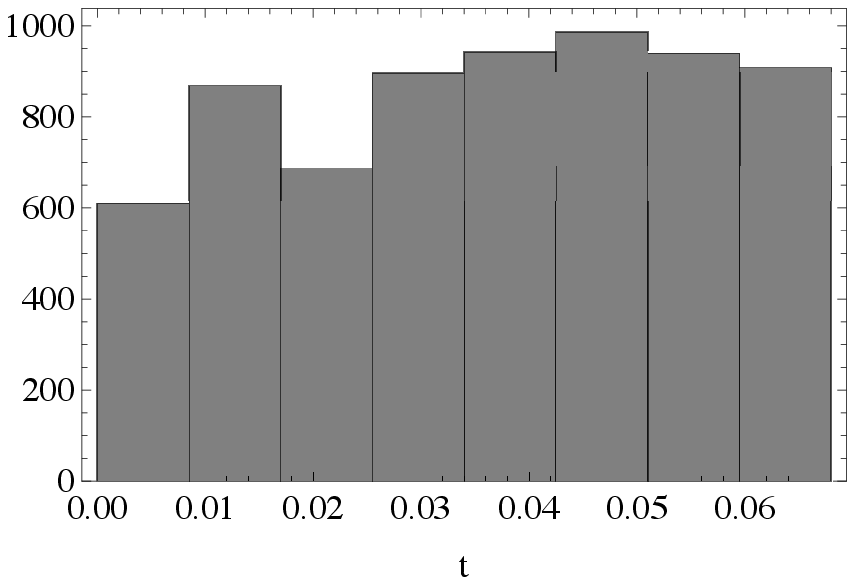} 
  \caption{
  \label{fig:t_histo}
  (Left) Final $\{\wt_\ell\}$ 
  and its polynomial fit $\wt(t)$. 
  (Right) Final histogram of $t$. 
  }
\end{figure}

We set $T_0=0$ while we choose $T_1$ depending on $\mu$ 
as in Table \ref{table:sim_parameters}. 
Other simulation parameters are also given in Table \ref{table:sim_parameters}. 
\begin{table}[ht]
  \centering
  \begin{small}
  \begin{tabular}{|c|c|c|c|c|c|c|}
    \hline
    $\mu$      & $0.4$  & $0.45 $ & $0.5$  & $0.55$ & $0.575$ & $0.6$  \\
    \hline
    $T_0$      &  0     &    0    &    0   &    0   &   0      &  0     \\
    $T_1$      &  0.025 &  0.048  &  0.056 &  0.068 & 0.068    &   0.068 \\
    $N_{\rm init}$
               &  80    &  60     &  40    &  100   & 100      &   50     \\
    $N_{\rm conf}$
               &  4,000 & 4,000   &  4,000 &  12,000& 10,000   &   6,000 \\

    $\hat{T}_0$ ($\langle\bar{\psi}\psi\rangle$)
               &  0.0145& 0.02784 & 0.028  & 0.04352 & 0.04216  & 0.0272\\
    $\hat{T}_1$ ($\langle\bar{\psi}\psi\rangle$)
               &  0.025  & 0.048 & 0.056  & 0.068  & 0.068    & 0.06664 \\
    $N_{\rm conf}(\hat{T}_0,\hat{T}_1)$ ($\langle\bar{\psi}\psi\rangle$)
               &  1,600  & 1,600   & 1,800   & 4,400   & 4,400     &  3,600   \\
    $\hat{T}_0$ ($\langle\psi^\dagger\psi\rangle$)
               &  0.0095&0.02112  & 0.03472& 0.04624& 0.03128  & 0.04352 \\
    $\hat{T}_1$ ($\langle\psi^\dagger\psi\rangle$)
               &  0.025  & 0.048 & 0.056  & 0.068  & 0.068    &  0.068\\
    $N_{\rm conf}(\hat{T}_0,\hat{T}_1)$ ($\langle\psi^\dagger\psi\rangle$)
               &  2,400  & 2,400    & 1,250   & 3,500   & 6,000    &  2,200  \\
    \hline
    $\mu$      & $0.625$& $0.65$  & $0.7$ & $0.75$ & $0.8$ & \\
    \hline
    $T_0$      &    0   &    0   &    0   &   0      & 0    &\\
    $T_1$      & 0.068  & 0.064   & 0.06  & 0.052  & 0.04   &\\
    $N_{\rm init}$
              &40      & 75      & 50    & 40     &  40    &\\
    $N_{\rm conf}$
              & 5,000  & 8,000   & 4,000 & 4,000  &  4,000  &\\

    $\hat{T}_0$ ($\langle\bar{\psi}\psi\rangle$)
               & 0.02312& 0.02432 & 0.0204& 0.01456&  0.0136&\\
    $\hat{T}_1$ ($\langle\bar{\psi}\psi\rangle$)
               & 0.068  & 0.064   & 0.06  & 0.052  &  0.04  &\\
    $N_{\rm conf}(\hat{T}_0,\hat{T}_1)$ ($\langle\bar{\psi}\psi\rangle$)
               & 3,500   & 5,500    & 2,600  & 2,500   & 2,600  &\\
    $\hat{T}_0$ ($\langle\psi^\dagger\psi\rangle$)
               &0.02448 &0.01664  &0.0252 &0.02496 &0.0264 &\\
    $\hat{T}_1$ ($\langle\psi^\dagger\psi\rangle$)
               & 0.068  & 0.064   &0.06   & 0.052  & 0.04 &\\
    $N_{\rm conf}(\hat{T}_0,\hat{T}_1)$ ($\langle\psi^\dagger\psi\rangle$)
               & 3,000   & 6,500    &2,500   & 1,750   & 1,200  & \\
    \hline
  \end{tabular}
  \end{small}
  \caption{Simulation parameters.}
  \label{table:sim_parameters}
\end{table}\noindent
There, 
$N_{\rm init}$ is the number of initial configurations, 
and $N_{\rm conf}$ is the number of configurations 
in the simulation range $[T_0,T_1]$, 
while $N_{\rm conf}(\hat{T}_0, \hat{T}_1)$ is 
that in the estimation range $[\hat{T}_0,\hat{T}_1]$, 
corresponding to the point $(\hat T_0,\hat T_1)$ chosen from a plateau. 
Note that $\hat T_0$ and $T_1$ depend on the choice of observables. 
We set the initial configuration to the final configuration 
in the test run determining $\wt(t)$. 
The tuning of $\wt(t)$ turns out to take two iterations  
to realize the condition \eqref{weight_convergence}. 
In Fig.~\ref{fig:t_histo}, 
we show the final form of $\wt(t)$ at $\mu=0.625$ 
and the resulting histogram of $t$. 
\noindent 

It sometimes happens that $\calR=\bigcup_t\Sigma_t$ 
is not well explored for large $t$ 
because of the complicated geometrical structure there. 
To facilitate transitions, 
at every start of the HMC algorithm 
we change the step size $\Delta s$ and the step number $n$ 
by randomly taking them 
from a set $\mathcal{C}=\{(\Delta_c,n_c\}$ $(c=1,\ldots,c_{\rm max})$.%
\footnote{
  This prescription is justified 
  by noticing that this gives a Markov chain 
  on an extended space $\calR\times\mathcal{C}$. 
} 
In the calculation below, 
we set $c_{\rm max}=3$ and choose $\mathcal{C}$ 
as in Table \ref{table:hmc_param}. 
\begin{table}[ht]
  \centering
  \begin{small}
  \begin{tabular}{|c|c|c|c|c|c|c|c|}
    \hline
    index $\idxq$     & $0$    & $1$    & $2$    & $3 $   \\
    \hline
    $\Delta s_\idxq$  &  0.01   &  0.005  & 0.001   & 0.00025   \\
    $n_\idxq$         &   25    &  50     &  50     &   100      \\
    \hline
  \end{tabular}
  \end{small}
  \caption{HMC parameters.}
  \label{table:hmc_param}
\end{table}\noindent

We comment that, 
if we use the original TLTM based on the parallel tempering, 
we need about 70 replicas for $n=10$. 
We list in Table \ref{table:nreplica} 
the numbers of replicas at $\mu=0.6$ for various $n$. 
\begin{table}[ht]
  \centering
  \begin{small}
  \begin{tabular}{|c|c|c|c|c|}
    \hline
    $n$        & 4      & 6       & 8      &  10    \\
    \hline
    $T$      &  0.    &  0.02   &  0.06  &  0.068  \\
    \#replicas  &  1    &  4      &  $\sim$\,33    &  $\sim$\,70   \\
    \hline
  \end{tabular}
  \end{small}
  \caption{Maximum flow time $T$ and the number of replicas at $\mu=0.6$.} 
  \label{table:nreplica}
\end{table}\noindent
Here, we first determine the maximum flow time $T$ 
so that the sign problem is well resolved there, 
and then determine the number of replicas 
so that the acceptance rate of the swapping is in the range 0.2--0.5.

\subsection{Results and analysis}
\label{sec:results}

\begin{figure}[ht]
\centering
  \includegraphics[width=70mm]{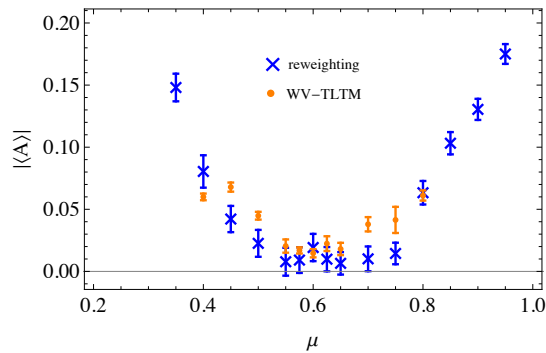}
  \caption{
  \label{fig:denominator}
  Average phase factors 
  $\langle A \rangle=\langle e^{-i\,{\rm Im}\,S(x)} \rangle_{\rm rewt}$ 
  from the naive reweighting method (blue) 
  and the average reweighting factors 
  $\langle A \rangle=\langle A(z) \rangle_\calR$ 
  from the WV-TLTM (orange). 
  The estimation range $[\hat T_0,\hat T_1]$ in the WV-TLTM 
  is set to that for the chiral condensate 
  (see Table \ref{table:sim_parameters}). 
}
\end{figure}

\begin{figure}[ht]
  \centering
  \includegraphics[width=70mm]{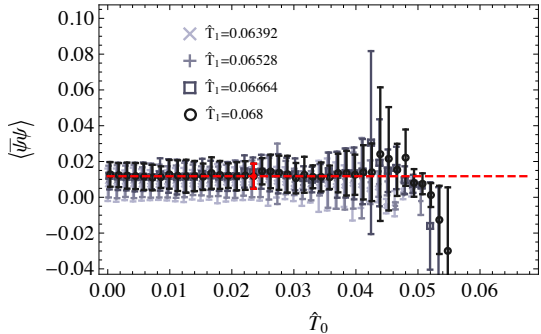} 
  \caption{
  \label{fig:plateau}
    Estimates of $\langle\bar\psi\psi\rangle$ at $\mu=0.625$ 
    with various $\hat T_0$ and $\hat T_1$. 
    The red filled circle is the point taken from a plateau 
    to be used for the estimation of $\langle\bar\psi\psi\rangle$, 
    and the horizontal dashed line is the estimate. 
  }
\end{figure}

\begin{figure}[ht]
\centering
  \includegraphics[width=70mm]{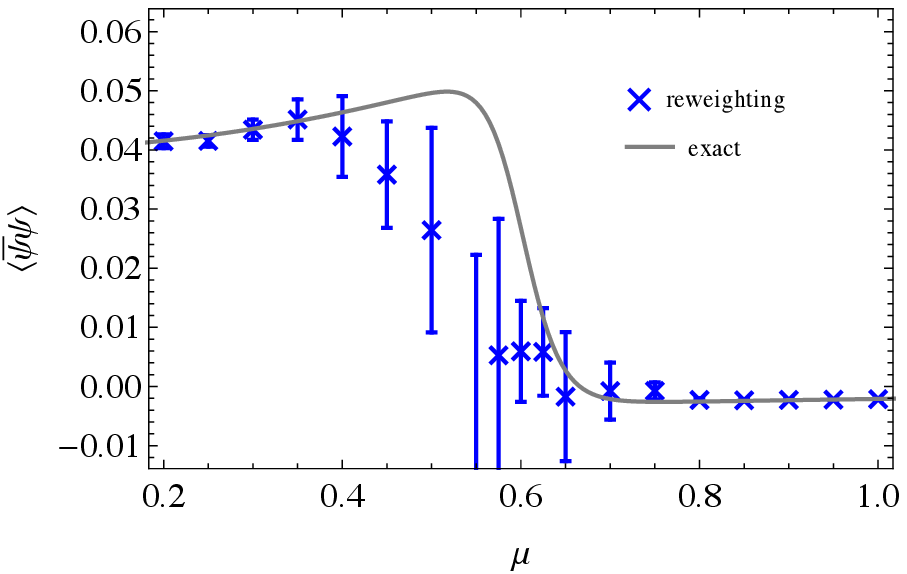} \hspace{5mm}
  \includegraphics[width=70mm]{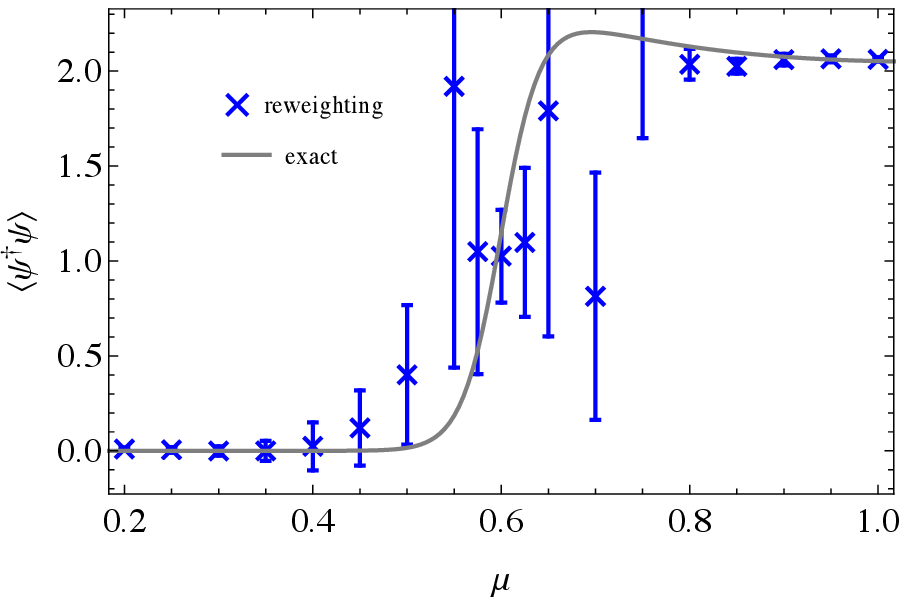} \\
  \includegraphics[width=70mm]{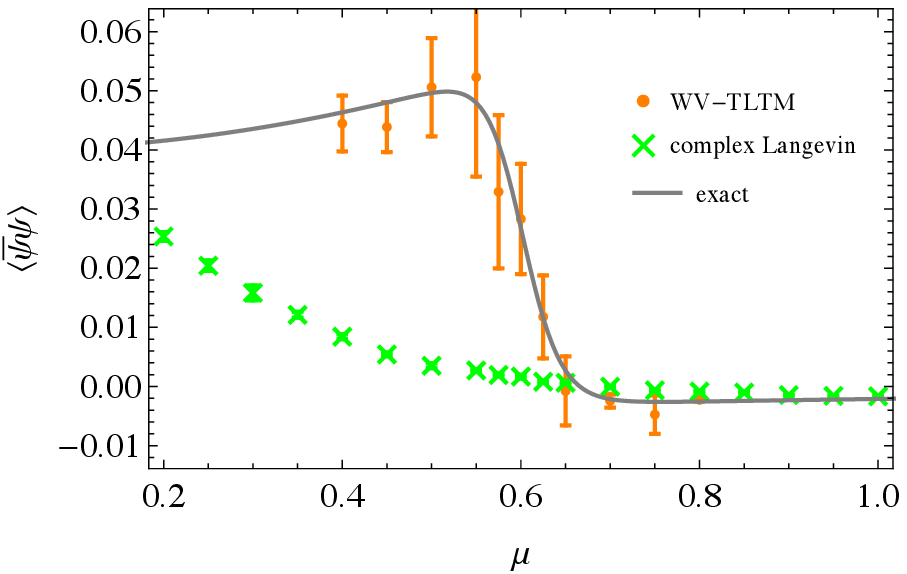} \hspace{5mm}
  \includegraphics[width=70mm]{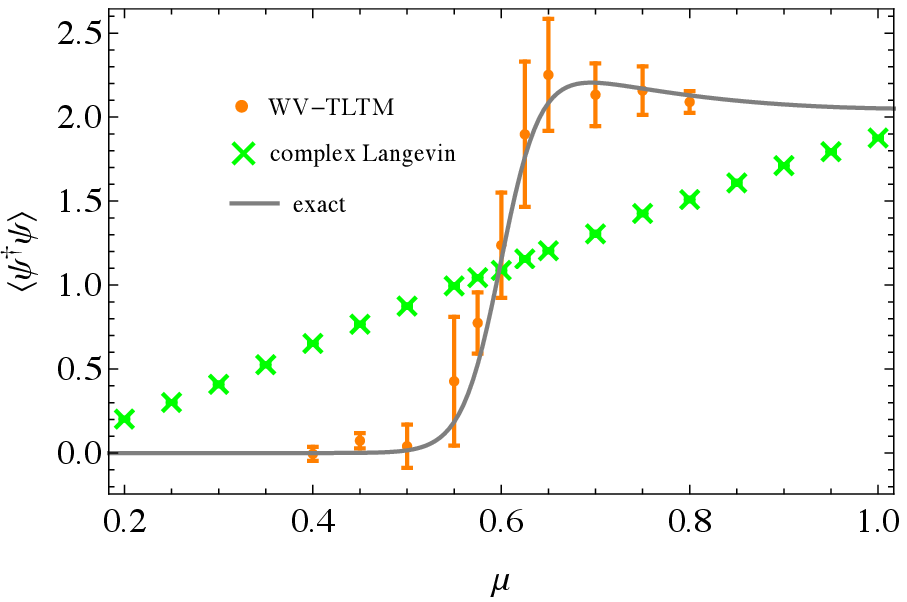} \\
  \caption{
  \label{fig:estimates}
  (Left) The chiral condensate. (Right) The Number density. 
  The top panels are the results from the reweighting method, 
  and the bottom panels are from the WV-TLTM and the complex Langevin method. 
}
\end{figure}

Figure \ref{fig:denominator} shows the average reweighting factors 
from the naive reweighting method (blue) 
and from the WV-TLTM (orange), 
the former exhibiting the existence of the sign problem around $\mu=0.6$. 
\noindent 
Figure \ref{fig:plateau} gives the estimates of $\langle\bar\psi\psi\rangle$
from the WV-TLTM 
at $\mu=0.625$ 
with various estimation ranges $[\hat T_0,\hat T_1]$. 
\noindent 
We see a plateau with a value close to the exact one 0.012.
Figure \ref{fig:estimates} exhibits 
the estimated values thus obtained 
for the chiral condensate $\langle\bar\psi\psi\rangle$ 
and the number density $\langle\psi^\dagger\psi\rangle$. 
\noindent 
As a comparison, 
we also display in the same figure 
the results from the naive reweighting method 
and the complex Langevin method, 
both with the sample size $N_{\rm conf}=10,000$. 
We see that 
the WV-TLTM correctly reproduces exact values, 
while the complex Langevin method suffers from the wrong convergence  
even for a parameter region free from the sign problem. 

\begin{figure}[ht]
  \centering
  \includegraphics[width=70mm]{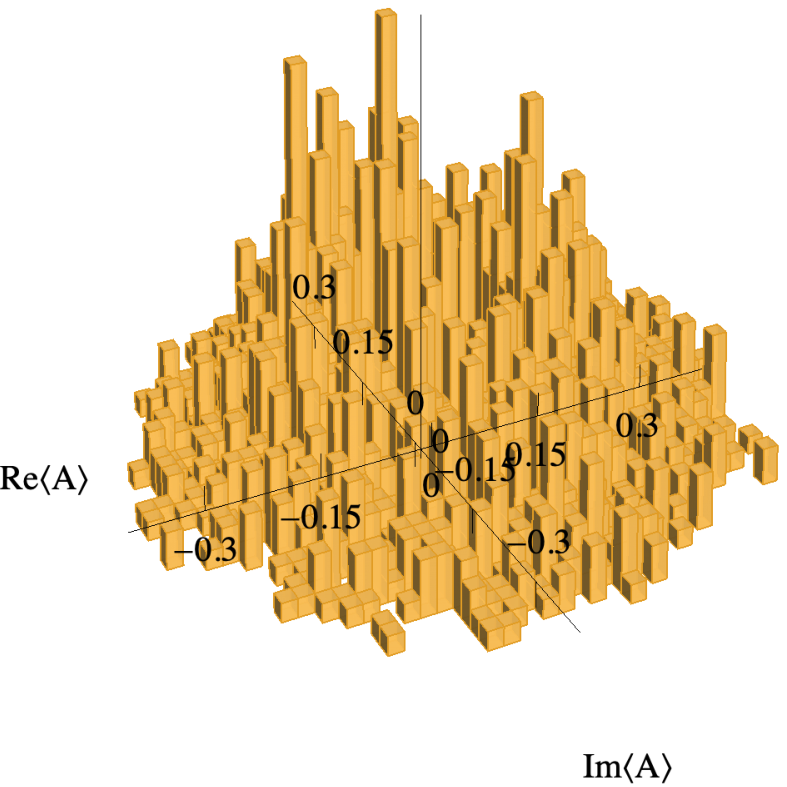} \hspace{5mm}
  \includegraphics[width=70mm]{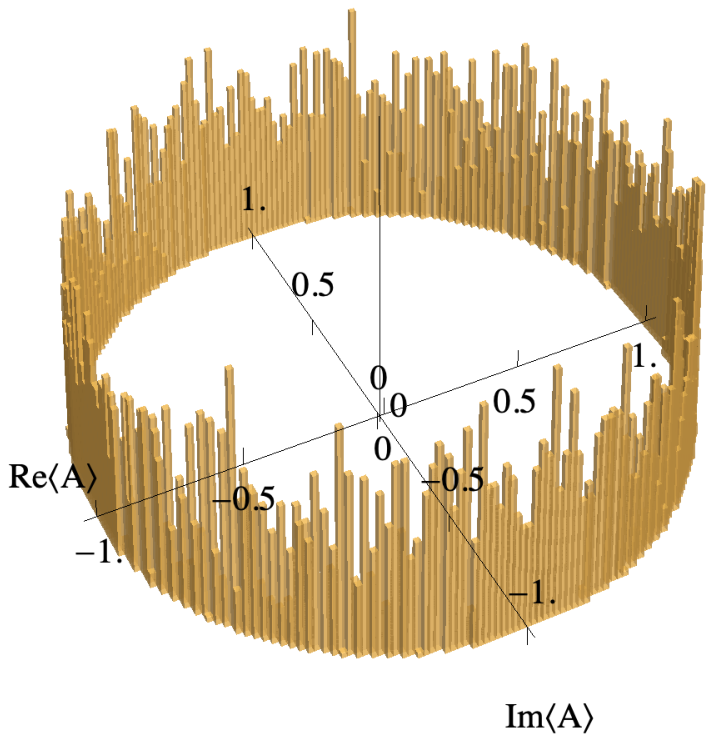}
  \caption{
  \label{fig:denom_distribution}
    (Left) Histogram of 
    $\denom(z)=\alpha^{-1}(z)\,e^{i\varphi(z)-i{\rm Im}\,S(z)}$ 
    obtained from the WV-TLTM. 
    (Right) Histogram of the phase factor 
    $e^{i\varphi(z)-i{\rm Im}\,S(z)}$ 
    obtained from the reweighting. 
    Both figures are at $\mu = 0.625$. 
  }
\end{figure}

\begin{figure}[ht]
  \centering
  \includegraphics[width=70mm]{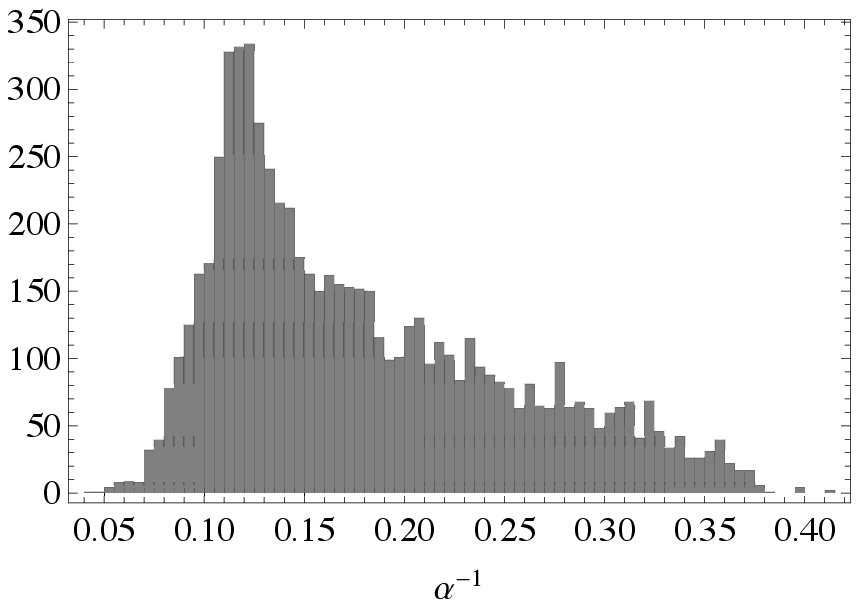} \hspace{5mm}
  \includegraphics[width=70mm]{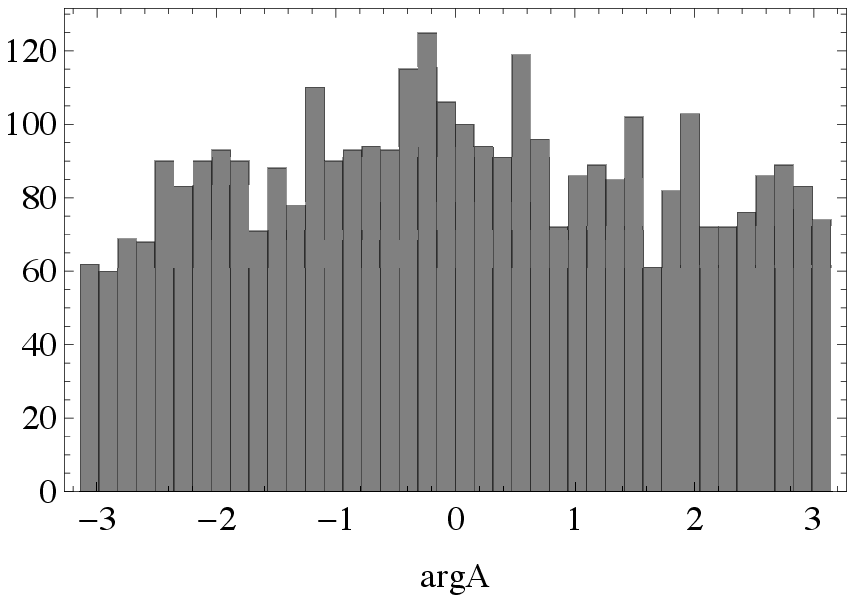}\\
  \includegraphics[width=70mm]{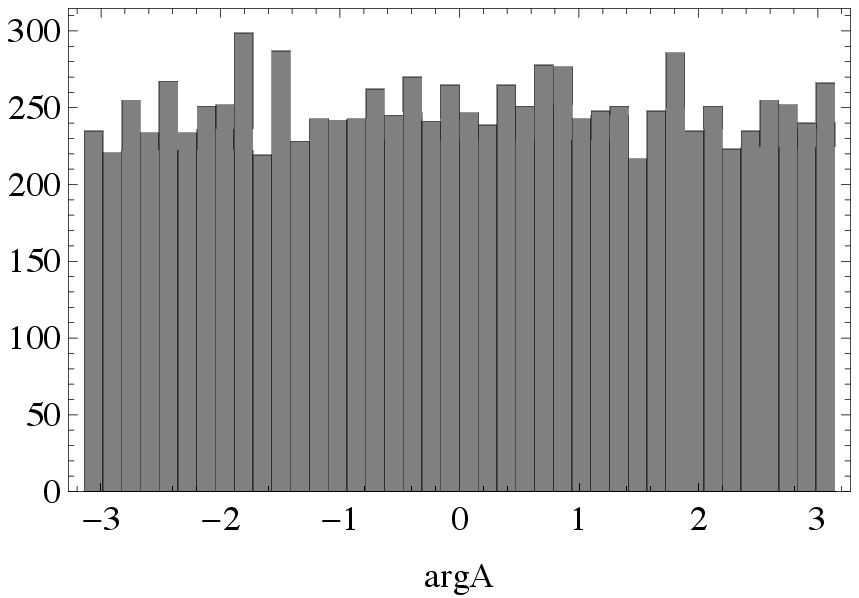}\\
  \caption{
  \label{fig:denom_distribution_rad_ang}
    (Top) Histograms of the modulus $|\denom(z)|=\alpha^{-1}(z)$ 
    and the phase ${\rm arg}\,\denom(z)=\varphi(z)-{\rm Im}\,S(z)$ 
    from the WV-TLTM. 
    (Bottom) Histogram of the phase from the reweighting method. 
  }
\end{figure}

One may find it strange that 
correct estimates are still obtained from the WV-TLTM 
even for such parameters that give small average reweighting factors 
$\langle \denom(z) \rangle_\calR$ (see Fig.\ref{fig:denominator}). 
To understand this, 
let us see Figs.~\ref{fig:denom_distribution} 
and \ref{fig:denom_distribution_rad_ang}, 
which show the histogram of 
$\denom(z)=\alpha^{-1}(z)\,e^{-i\,{\rm Im}\,S(z) + i \varphi(z)}$, 
and those of its modulus and phase, 
in the estimation range $[\hat{T}_0,\hat{T}_1]$ at $\mu=0.625$. 
\noindent 
\noindent 
We observe 
that the distribution of $\alpha^{-1}(z)$ has a peak 
around $\alpha^{-1}(z)=0.12$. 
The point is that this small value 
reduces not only the mean value of $\denom$, 
but also the statistical errors. 
This is in sharp contrast to the situation in the naive reweighting 
(see the right panel of Fig.~\ref{fig:denom_distribution}). 
In fact, 
in the latter (the naive reweighting), 
the reweighting factor is actually a phase factor, 
and is distributed uniformly on a unit circle, 
giving a vanishingly small average phase factor. 
The statistical errors are then of $O(1/\sqrt{N_{\rm con}})$,
because the standard deviation of phase factors 
for uniformly distributed phases 
is of $O(1)$. 
One thus needs a huge sample size of $e^{O(N)}$ 
to make the statistical errors relatively small to the mean value. 
In the WV-TLTM, in contrast, 
the reweighting factor $\denom(z)$ is distributed 
in a two-dimensional way (not on a circle), 
and the contributions of the radius $\alpha^{-1}$ enter 
both the mean value and the statistical errors, 
and also both the numerator and the denominator.  
Thus, the effect of small radius cancels out in a ratio of reweighted averages. 
Therefore, no additional problem is caused 
by the smallness of the reweighting factor, 
and the extent of the sign problem is still governed by the phase factor, 
which is reduced by taking sufficiently large flow times.

\section{Conclusion and outlook}
\label{sec:conclusion}

In this paper, 
we proposed a HMC algorithm 
on the worldvolume $\calR$ of an integration surface $\Sigma_t$,
where the flow time $t$ changes 
in the course of molecular dynamics, 
and thus the multimodal problem is resolved 
without introducing replicas. 
Furthermore, the computation of the Jacobian is not necessary 
in generating a configuration. 
We applied this algorithm to 
a chiral random matrix model (the Stephanov model) 
and confirmed that it reproduces correct results, 
solving both sign and multimodal problems simultaneously. 

The validity of this algorithm should be further investigated 
by applying it to other systems 
that also have the sign problem, 
including finite density QCD, strongly correlated  electron systems 
and real-time quantum field theories 
as well as frustrated spin systems 
like the antiferromagnetic Heisenberg model on the triangular lattice 
and the Kitaev model on the honeycomb lattice. 

It is important to keep developing the algorithm as well 
in order to perform large scale calculations 
for such systems listed above with less computational cost.  
For example, it should be nice 
if we can find a more efficient algorithm to tune $\wt(t)$.%
\footnote{
  Machine learning may be one of the possible tools. 
} 
At the same time, it is worth developing an algorithm 
where the weight $e^{-\wt(t)}$ needs not be introduced, 
as what happens when one switches from the simulated tempering 
\cite{Marinari:1992qd}
to the parallel tempering 
\cite{Swendsen1986,Geyer1991,Hukushima1996}. 
It is also desirable 
if one can construct an algorithm 
to evaluate the phase $e^{i\varphi}=\det J/|\det J|$ 
without computing the matrix elements of $J$ explicitly. 
Furthermore, 
in order to make a more accurate statistical analysis, 
it is important to develop a systematic method 
to estimate numerical errors 
that are necessarily introduced 
in integrating the antiholomorphic gradient flow 
and in solving Newton's method iteratively 
(Step 2 in subsection \ref{sec:summary}).

The modification of the flow equation \eqref{flow_zC} 
should also deserve intensive investigation for various reasons. 
To see this, 
note that  \eqref{flow_zC} is not the only possible equation 
deforming the original integration surface $\bbR^N$ 
so as to approach a union of Lefschetz thimbles. 
For example, it can be modified 
with a positive hermitian matrix $G^{ij}(z,z^\ast)$
to the form  
\begin{align}
  \frac{dz_t^i}{dt} &= G^{ij}(z,z^\ast)\,[\partial_j S(z_t)]^\ast,
\label{flow_zC2}
\end{align}
without changing the structure of thimbles. 
However, this modification changes flows of configurations off the thimbles, 
and can be designed 
so that flowed configurations approach zeros  of $e^{-S(z)}$ only very slowly 
(see, e.g., \cite{Tanizaki:2017yow}). 
We have investigated this type of modification, 
proposing to take $G^{ij}$ of the following simple form \cite{FMU_JPS2020}:
\begin{align}
  G^{ij}(z,z^\ast)=\frac{\delta^{ij}}{1+(|\partial S(z)|/\Lambda)^{\alpha}}
  \quad (\alpha\geq 2). 
\end{align}
This actually removes zeros from $\calR$ for finite flow times, 
and sometimes is helpful 
in iteratively solving the constraint \eqref{constraint_zprime2}. 
However, it seems that 
the obtained gain does not exceed the increased complexity 
of the algorithm, 
and also that 
the functional form of $G^{ij}$ needs to be fine-tuned, 
depending on parameters on each model. 
This is the reason why we did not pursue this possibility in this paper. 
However, it will be essentially important 
when one develops a Metropolis-Hastings algorithm 
described in appendix \ref{sec:metropolis}, 
because the configuration space 
$\tilde{\calR}=\{\xi=(t,x^a)\,|\,T_0\leq t\leq T_1\}$ 
comes to have a simple structure 
if points to be mapped to zeros do not exist in the region. 

Another possible application of modifying the flow equation 
is to provide a mechanism to solve the so-called global sign problem 
(cancellation among contributions from different thimbles). 
In fact, 
since $\alpha^{-1}(z)$ increases exponentially 
in the vicinity of a Lefschetz thimble 
[see a comment below \eqref{lapse_growth}], 
a change of flows caused by the modification 
may significantly shift the distribution of $\denom(z)$ 
and distort the balanced contributions from different thimbles, 
that was the origin of the global sign problem. 

A study along these lines is now in progress and will be reported elsewhere.

\section*{Acknowledgments}
The authors thank 
Hitotsugu Fujii, Issaku Kanamori, Yoshio Kikukawa, 
Yusuke Namekawa, and especially Naoya Umeda 
for useful discussions. 
This work was partially supported by JSPS KAKENHI 
(Grant Numbers 20H01900, 18J22698)   
and by SPIRITS 2020 of Kyoto University (PI: M.F.).

\appendix

\section{Geometry of the RATTLE algorithm}
\label{sec:RATTLE_geometry}

In this appendix, 
we clarify the geometrical aspects of the RATTLE algorithm 
and prove a few statements necessary for discussions in the main text.

As in the main text, 
let $\calR$ be an $m$-dimensional manifold 
embedded in the flat space $\bbR^M=\{z=(z^I)\}$ $(I=1,\ldots,M)$.%
\footnote{
  When $\calR$ is the worldvolume of an integration surface 
  in $\bbC^N=\bbR^{2N}$,
  we set $M=2N$ and $m=N+1$. 
} 
With local coordinates $\xi=(\xi^\mu)$ $(\mu=1,\ldots,m)$ of $\calR$, 
the embedding is expressed by functions $z^I=z^I(\xi)$. 
The vectors
\begin{align}
  E_\mu = \bigl(E^I_\mu=\partial z^I/\partial \xi^\mu\bigr) 
\end{align}
form a basis of the tangent space $T_z\calR$ at $z\in\calR$, 
and give the induced metric $ds^2=g_{\mu\nu}(\xi)\,d\xi^\mu d\xi^\nu$ 
as  
\begin{align}
  g_{\mu\nu} = E_\mu \cdot E_\nu.
\end{align}
Furthermore, 
denoting the inverse of $(g_{\mu\nu})$ by $(g^{\mu\nu})$, 
and defining another basis of $T_z\calR$ by%
\footnote{
  Note that $E^\mu\cdot E^\nu=g^{\mu\nu}$. 
} 
\begin{align}
  E^\mu \equiv g^{\mu\nu} E_\mu, 
\end{align}
we also introduce local coordinates $\eta=(\eta_\mu)$ on $T_z\calR$ 
as coefficients in an expansion with respect to $E^\mu$.%
\footnote{
  As in the main text, 
  we denote a function on $\calR$ by $f(z)$ and $f(\xi)$, 
  interchangeably, 
  with the understanding that $z=z(\xi)$. 
  The transition matrix is also written 
  as $P(z'|z)$ and $P(\xi'|\xi)$ 
  for $z=z(\xi),\,z'=z(\xi')\in\calR$.
  Similarly, a function on $T\calR$  
  is written as $f(z,\pi)$ and $f(\xi,\eta)$, interchangeably. 
} 
Namely, for $\pi\in T_z\calR$, 
its coordinates $\eta=(\eta_\mu)$ are defined through the relation   
\begin{align}
  \pi = \eta_\mu\, E^\mu,
\end{align}
whose explicit forms are given by
\begin{align}
  \eta_\mu = E_\mu\cdot\pi.
\end{align}
The line element on $T_z\calR$ then takes the form 
\begin{align}
  (d\pi^I)^2\bigr|_{T_z\calR} = g^{\mu\nu}(\xi)\, d\eta_\mu\, d\eta_\nu,
\end{align}
and thus the volume elements of $T\calR$ 
is given by 
\begin{align}
  dV_{T\calR} = Dz\, D\pi = D\xi\, D\eta
\label{phase-space_vol}
\end{align}
with 
\begin{align}
  Dz \equiv D\xi \equiv \sqrt{g}\,d\xi, \quad
  D\pi \equiv D\eta \equiv d\eta/\sqrt{g}. 
\end{align}

We think of the tangent bundle $T\bbR^M=\{(z^I,\pi^I)\}$ 
({\em not} the cotangent bundle) 
as the phase space of motions in $\bbR^M$ 
with a symplectic form $\Omega\equiv d\pi^I\wedge dz^I$. 
The sub-bundle 
$T\calR=\{(z^I,\pi^I)\,|\,z\in\calR,\,\pi\in T_z\calR\}$ 
is then regarded as the phase space of constrained motions on $\calR$. 
Its symplectic form is given by%
\footnote{
  This can be proved as follows: 
  $\omega=d\pi^I\wedge dz^I|_{T\calR}
  =\bigl[d\pi^I|_{T_z\calR}\bigr]\wedge dz^I(\xi)
  =\bigl[d\pi^I|_{T_z\calR}\bigr]\wedge E^I_\mu(\xi)\,d\xi^\mu
  = d\bigl[\pi^I|_{T_z\calR}\,E^I_\mu(\xi)\bigr]\wedge d\xi^\mu
  = d\eta_\mu\wedge d\xi^\mu$,
  where we have used $d\bigl[E^I_\mu(\xi)\bigr]\wedge d\xi^\mu
  =\bigl(\partial^2 z^I/\partial\xi^\nu\partial\xi^\mu\bigr)\,
  d\xi^\nu\wedge d\xi^\mu=0$.
} 
\begin{align}
  \omega \equiv \Omega|_{T\calR} = d\eta_\mu\wedge d\xi^\mu,
\end{align}
which defines the Poisson bracket as 
$\{\xi^\mu,\eta_\nu\}=\delta^\mu_\nu$, 
$\{\xi^\mu,\xi^\nu\}=\{\eta_\mu,\eta_\nu\}=0$. 
The volume element \eqref{phase-space_vol} 
agrees with the phase-space volume element 
associated with $\omega$:
\begin{align}
  dV_{T\calR} =\frac{\omega^m}{m!},
\label{phase-space_vol2}
\end{align}
because 
$\omega^m/m!=d\xi\,d\eta=(\sqrt{g}\,d\xi)\,(d\eta/\sqrt{g})=D\xi\,D\eta$.

We now consider molecular dynamics on $T\calR$ with a Hamiltonian 
\begin{align}
  H(\xi,\eta) = \frac{1}{2}\,(\pi^I)^2 + V(z) \Bigr|_{T\calR}
  = \frac{1}{2}\,g^{\mu\nu}(\xi)\,\eta_\mu \eta_\nu + V(z(\xi)), 
\end{align}
with which the time evolution of $(\xi,\eta)$ is given by
\begin{align}
  \partial_s \xi^\mu = \{\xi^\mu,H\} = \partial_{\eta_\mu}H(\xi,\eta),
  \quad
  \partial_s \eta_\mu = \{\eta_\mu,H\} = \partial_{\xi^\mu}H(\xi,\eta).
\label{Hamilton_eq}
\end{align}
It is easy to see that 
this evolution preserves  
the Hamiltonian $H$, the symplectic form $\omega$, 
and thus also the phase-space volume element $dV_{T\calR}$: 
\begin{align}
  \partial_s H = 0, \quad \partial_s \omega=0, \quad 
  \partial_s (dV_{T\calR}) = 0.
\end{align}
We write the motion from $(\xi,\eta)$ to $(\xi',\eta')$ 
with time interval $s$ as a map $\Phi_s$:
\begin{align}
  (\xi,\eta)\to(\xi',\eta') = \Phi_s(\xi,\eta)
  \equiv (\xi_s(\xi,\eta),\eta_s(\xi,\eta)).
\end{align}
Since the Hamiltonian satisfies the relation 
$H(\xi,-\eta)=H(\xi,\eta)$, 
the map $\Phi_s$ preserves the reversibility. 
Namely, if $(\xi',\eta')=\Phi_s(\xi,\eta)$ is a motion, 
we have $(\xi,-\eta)=\Phi_s(\xi',-\eta')$.
Furthermore, due to the volume preservation, 
the kernel 
\begin{align}
  \Phi_s(\xi',\eta'|\xi,\eta) \equiv 
  \delta\bigl(\xi'-\xi_s(\xi,\eta)\bigr)\,
  \delta\bigl(\eta'-\eta_s(\xi,\eta)\bigr)
\end{align}
satisfies the relation 
\begin{align}
  \Phi_s(\xi',\eta'|\xi,\eta) = \Phi_s(\xi,-\eta|\xi',-\eta').
\end{align}

For stochastic processes on $\calR$, 
we define the probability density $p(z(\xi))=p(\xi)$ on $\calR$ 
with respect to 
the volume element $\muR=D\xi=\sqrt{g}\,d\xi$, 
which is thus normalized as 
\begin{align}
  \int_\calR D\xi\,p(\xi) = 1.
\end{align}
One then can easily show that the transition matrix 
\begin{align}
  P_s(\xi'|\xi) \equiv 
  \int_{T_{z'}\calR} D\eta' \int_{T_z\calR} D\eta\,
  \Phi_s(\xi',\eta'|\xi,\eta)\,\frac{1}{(2\pi)^{m/2}}\,
  e^{-g^{\mu\nu}(\xi)\,\eta_\mu\eta_\nu/2}
\label{trans_mat}
\end{align}
satisfies the detailed balance condition%
\footnote{
  This will be proved for a more complicated case 
  in \eqref{detailed_balance_proof}.
} 
\begin{align}
  P_s(\xi'|\xi)\,e^{-V(z(\xi))} =  P_s(\xi|\xi')\,e^{-V(z(\xi'))} 
\end{align}
and the normalization condition 
\begin{align}
  \int_\calR D\xi'\,P_s(\xi'|\xi)\ = 1.
\end{align}

The Gaussian distribution 
$e^{-g^{\mu\nu}(\xi)\,\eta_\mu\eta_\nu/2}/(2\pi)^{m/2}$ 
in \eqref{trans_mat} can be obtained 
by first generating $\tilde{\pi}=(\tilde{\pi}^I)\in T_z\bbR^M$  
from the Gaussian distribution 
$e^{-\tilde\pi^2/2}/(2\pi)^{M/2}$
and then projecting it onto $T_z\calR$. 
In fact, 
for the orthogonal decomposition 
$\tilde{\pi} = \pi + \pi_\perp$, 
$\tilde\pi^2$ is written as $\tilde\pi^2=\pi^2+\pi_\perp^2$, 
and the integration measure of $T_z \bbR^M$ is factorized
as 
$d^{M}\tilde\pi\equiv d\tilde{\pi}^1 \cdots d\tilde{\pi}^{M} 
= \muTR\, D\pi_\perp$. 
By integrating out only the normal component, 
the distribution of $\pi \in T_z\calR$ is left 
in the desired form: 
\begin{align}
  \int_{N_z\calR} d^{M}\tilde\pi\, 
  \frac{1}{(2\pi)^{M/2}} e^{-\tilde{\pi}^2/2 } 
  = \muTR\,\int_{N_z\calR} D\pi_\perp \, \frac{1}{(2\pi)^{M/2}}\,
  e^{-(\pi^2 + \pi_\perp^2)/2 } 
  = \frac{1}{(2\pi)^{m/2}}\, e^{-\pi^2/2 }\,\muTR. 
\end{align}
Note that the projector $\Pi_\calR$ 
from $\tilde\pi\in T_z\bbR^M$ 
to $\pi=\Pi_\calR\,\tilde\pi \in T_z\calR$ is given by 
\begin{align}
  \Pi_\calR = E_\mu (E^\mu)^T = g^{\mu\nu} E_\mu E_\nu^T.
\end{align}

We now assume that $\calR$ is characterized by $M-m$ independent 
constraint equations, $\phi^r(z)=0$ $(r=1,\ldots,M-m)$. 
Then, Hamilton's equations \eqref{Hamilton_eq} 
can be expressed as constrained motions in $\bbR^M$ 
by using Lagrange multipliers $\lambda_r$: 
\begin{align}
  \partial_s z &= \pi,
\label{zdot_app}
\\
  \partial_s \pi &= -\partial V(z) - \lambda_r\, \partial \phi^r(z),
\label{pidot_app}
\\
  \phi^r(z) &= 0,
\label{constraint_z_app}
\\
  \pi \cdot \partial \phi^r(z) &= 0.
\label{constraint_pi_app} 
\end{align}
Here, $\partial \equiv (\partial_{z^I})$ is the gradient in $\bbR^M$. 

The RATTLE algorithm \cite{Andersen:1983,Leimkuhler:1994} 
is an algorithm 
which discretizes \eqref{zdot_app}--\eqref{constraint_pi_app} 
preserving the symplecticity and the reversibility 
(below $\Delta s$ is the step size): 
\begin{align}
  \pi_{1/2} &= \pi - \frac{\Delta s}{2}\, \partial V(z) 
  - \lambda_r\, \partial \phi^r(z),
\label{pihalf_app}
\\
  z' &= z + \Delta s \, \pi_{1/2},
\label{zprime_app}
\\
  \pi' &= \pi - \frac{\Delta s}{2}\, \partial V(z') 
  - {\lambda'}_r\, \partial \phi^r(z').
\label{piprime_app}
\end{align}
Here, $\lambda_r$ and $\lambda'_r$ are determined, respectively, 
so that the following constraints are satisfied: 
\begin{align}
  z' &\in \calR \quad (\mbox{i.e.~}\phi^r(z')=0), 
\label{constraint_zprime_app}
\\
  \pi' &\in T_{z'} \calR.
\label{constraint_piprime_app}
\end{align}
One can easily show that 
the map $\Phi_{\Delta s}:\,(z,\pi) \rightarrow (z',\pi')$ 
actually satisfies the symplecticity and the reversibility 
(with $\lambda_r$ and $\lambda_r'$ interchanged): 
\begin{align}
  &\bullet~\omega(z',\pi') = \omega(z,\pi), 
\\
  &\bullet~ (z',\pi') = \Phi_{\Delta s}(z,\pi) 
  \Rightarrow (z, -\pi) = \Phi_{\Delta s}(z',-\pi'), 
\end{align}
which means that 
\begin{align}
  \Phi_{\Delta s}(z',\pi'|z,\pi) = \Phi_{\Delta s}(z,-\pi|z',-\pi'). 
\label{sympl_rev_rattle}
\end{align}
The Hamiltonian is conserved to the order of $\Delta s^2$, 
i.e., $H(z',\pi')-H(z,\pi) = O(\Delta s^3)$. 

The HMC algorithm then consists of the following three steps 
for a given initial configuration $z\in\calR$: 
\begin{description}

\item[Step 1.]
  Generate a vector $\tilde\pi=(\tilde{\pi}^I)\in T_z\bbR^M$ 
  from the Gaussian distribution 
  \begin{align}
    \frac{1}{(2\pi)^{M/2}} e^{- \tilde{\pi}^2/2 }, 
  \end{align}
  and project it onto $T_z\calR$ 
  to obtain an initial momentum $\pi=(\pi^I)\in T_z\calR$.

\item[Step 2.]
  Calculate $\Phi_{\Delta s}(z,\pi)$ 
  from \eqref{pihalf_app}--\eqref{constraint_piprime_app}. 
  We repeat this step $n$ times 
  to obtain $(z',\pi') = \Phi_{\Delta s}^n(z,\pi)$ 

\item[Step 3.]
  Update the configuration $z$ to $z'$ with a probability 
  \begin{align}
    \min\big( 1, e^{-H(z',\pi') + H(z,\pi)} \big). 
  \label{acceptance_probability}
  \end{align}
\end{description}

The above process defines a stochastic process on $\calR$ 
with the following transition matrix for $z'\neq z$: 
\begin{align}
  P(z'|z) \equiv \int_{T_{z'}\calR} \muTRp \,
  \int_{T_z\calR} \muTR \,
  \min\big( 1, e^{-H(z',\pi') + H(z,\pi)} \big)\,
  \Phi_{\Delta s}^n(z',\pi'|z,\pi)\,
  \frac{ e^{-\pi^2/2} }{(2\pi)^{m/2}}.
\label{eq:transition_matrix}
\end{align}
The diagonal ($z'=z$) components are determined 
from the probability conservation. 
$P(z'|z)$ can be shown to satisfy the detailed balance condition 
\begin{align}
  P(z'|z)\,e^{-V(z)} = P(z|z')\,e^{-V(z')} \quad (z,\,z'\in\calR)
\end{align}
as follows: 
\begin{align}
  &P(z'|z)\,e^{-V(z)} = \int_{T_{z'}\calR} \muTRp \,
  \int_{T_z\calR} \muTR \,
  \min\big( 1, e^{-H(z',\pi') + H(z,\pi)} \big)\,
  \Phi_{\Delta s}^n(z',\pi'|z,\pi)\,
  \frac{ e^{-H(z,\pi)} }{(2\pi)^{m/2}}
\nonumber
\\
  &= \frac{1}{(2\pi)^{m/2}}\,
  \int_{T_{z'}\calR} \muTRp \,
  \int_{T_z\calR} \muTR \,
  \min\big( e^{-H(z,\pi)}, e^{-H(z',\pi')} \big)\,
  \Phi_{\Delta s}^n(z,-\pi|z',-\pi')
\nonumber
\\
  &= P(z|z')\,e^{-V(z')},
\label{detailed_balance_proof}
\end{align}
where we have used \eqref{sympl_rev_rattle} $n$ times 
to get the second line, 
and have made the change of integration variables, 
$\pi\to -\pi$ and $\pi'\to -\pi'$, 
with the relation $H(z,-\pi)=H(z,\pi)$ 
to obtain the third line.

\section{Analytical expressions for the Gaussian case}
\label{sec:gaussian} 

We present the analytical expressions 
for some geometrical quantities 
defined in subsection \ref{sec:wv_geometry} 
for the action 
\begin{align}
  S(x) = \frac{\beta}{2}\sum_{k=1}^N (x^k - i)^2. 
\end{align}
This has a single critical point 
at $z_\sigma=(z_\sigma^k=i)$, 
and the corresponding Lefschetz thimble 
is given by 
$\mathcal{J}_\sigma=\{z=(z^k)\in\bbC^k\,|\,{\rm Im}\,z^k=1~(\forall k)\}$. 
Complex vectors will be used throughout this appendix. 

The solution of the antiholomorphic flow equation \eqref{flow_zC} 
takes the form 
\begin{align}
  z^k(t,x) = x^k e^{\beta t} + i (1-e^{-\beta t}). 
\end{align}
The Jacobian $J(t,x)$ is thus given by 
\begin{align}
  {J^k}_a(t,x) = \frac{\partial z^k(t,x)}{\partial x^a} 
  = e^{\beta t}\, \delta^k_a. 
\end{align}
The tangent vectors $E_0$ and $E_a$ are 
\begin{align}
  E_0^k = \beta\, (z^k - i)^\ast = \beta\,(x^k e^{\beta t}+i e^{-\beta t}), 
  \quad 
  E_a^k = e^{\beta t}\, \delta^k_a. 
\end{align}
Using \eqref{E0perp_C}, we have
\begin{align}
  (E_0^\perp)^k = -i\beta\,({\rm Im}\, z^k - 1) = i \beta\, e^{-\beta t}. 
\end{align}
The components of $g_{\mu\nu}$ are then given 
in the ADM parametrization \eqref{ADM} by 
\begin{align}
  \gamma_{ab} &= (J^\dagger J)_{ab} = e^{2\beta t}\,\delta_{ab}, 
\\
  \beta^a &= \gamma^{ab}\, {\rm Re}\,(E_0^\dagger E_b) = \beta\, x^a, 
\\
  \alpha &= |E_0^\perp| = \beta \sqrt{N}\, e^{-\beta t}. 
\label{lapse_growth}
\end{align}
We see that the inverse lapse is given 
by $\alpha^{-1}=e^{\beta t}/(\beta\sqrt{N})$ 
and increases exponentially in flow time $t$ 
as $z(t,x)$ approaches the Lefschetz thimble. 

The ideal weight function $e^{-\wt(t)}$ 
giving a uniform distribution of $t$ 
is given by [see \eqref{ideal_weight}] 
\begin{align}
  e^{+W(t)} = 
  \int_{\bbR^N} dx \,
  \lapse(t,x)\, |\det J(t,x)|\, e^{-{\rm Re}\,S(z(t,x))}
  = \beta \sqrt{N}\, \Big(\frac{2\pi}{\beta}\Big)^{N/2}\, 
  e^{-\beta t +(\beta N/2)e^{-2\beta t}}, 
\end{align}
and thus we have 
\begin{align}
  \wt(t) = -\beta t + \frac{\beta N}{2}\,e^{-2 \beta t}.
\end{align}
We have ignored $t$-independent constants. 
We see that the weight factor also increases exponentially, 
$e^{-W(t)}\simeq e^{\beta t}$, 
at large flow times.

\section{Proof of eq.~\eqref{grad_t_general}}
\label{sec:grad_t_proof} 

In this appendix, we prove the equality \eqref{grad_t_general}. 
First, we construct $2N$ coordinates $(\zeta^A) \equiv (\xi^\mu, \phi^r)$ 
in the vicinity of $\calR$ in $\bbR^{2N}$, 
by regarding $(\phi^r)$ as coordinates in the extra dimensions, 
and introduce at each point $z\in\calR$ 
a basis $\{E_A\}$ of the tangent space $T_z\bbR^{2N}$ as 
\begin{align}
  E_A = \bigl(E_A^I \equiv \partial z^I / \partial \zeta^A \bigr). 
\end{align}
We further introduce the dual basis $\{\hat{E}^A\}$ to $\{E_A\}$ by 
\begin{align}
  \hat{E}^A = \bigl(\hat{E}^{A\,I} \equiv 
  \partial \zeta^A / \partial z^I\bigr), 
\end{align}
which satisfy 
\begin{align}
  \hat{E}^A \cdot E_B = \delta^A_B. 
\label{dual}
\end{align}
Note that $\hat{E}^0$ and $\hat{E}^r$ 
equal the gradients $\partial t(z)$ and $\partial\phi^r(z)$, 
respectively. 
Then, since the vectors $E_0^\perp$, $\hat{E}^a$ and $\hat{E}^r$ 
also form a basis of $T_z\bbR^{2N}$, 
$\hat{E}^0$ can be expanded in the form 
\begin{align}
  \hat{E}^0 = c_\perp E_0^\perp + c_a \hat{E}^a + c_r \hat{E}^r.  
\end{align}
The coefficients can be calculated 
by using the relations \eqref{dual} 
and $E_0\cdot E_0^\perp=(E_0^\perp)^2$
to be 
\begin{align}
  c_\perp = \frac{1}{(E_0^\perp)^2},\quad
  c_a = 0, \quad
  c_r = -(E_r\cdot E_0^\perp)\times c_\perp 
  = -\frac{E_r\cdot E_0^\perp}{(E_0^\perp)^2}, 
\end{align}
and thus we find that  $\hat{E}^0=\partial t(z)$ takes the form 
\begin{align}
  \partial t(z) = \frac{1}{(E_0^\perp)^2}\, E_0^\perp 
  -\frac{E_r\cdot E_0^\perp}{(E_0^\perp)^2}\, \partial\phi^r(z). 
\end{align}

\section{Another version of WV-TLTM with Metropolis-Hastings algorithm}
\label{sec:metropolis} 

In this appendix, 
we give another version of WV-TLTM, 
which also does not require the computation of the Jacobian 
in generating a configuration. 
This is a Metropolis-Hastings algorithm 
on a subspace in the parameter space (not in the target space), 
$\tilde\calR\equiv\{\xi=(t,x^a)\,|\,T_0\leq t\leq T_1\}$. 

We first rewrite \eqref{integral_R_intro} to the form
\begin{align}
  \langle \mathcal{O}(x) \rangle
  &= \frac{\int d\xi \,
    \det J(\xi)\, e^{- S(z(\xi))-\wt(t)}\, \mathcal{O}(z(\xi)) }
    {\int d \xi \,
    \det J(\xi)\, e^{- S(z(\xi)) -\wt(t)} },
\label{integral_R2}
\end{align}
where $d\xi\equiv dt\,dx \equiv dt\,dx^1\cdots dx^N$. 
Then, by introducing a new positive weight and a new reweighting factor 
as 
\begin{align}
  e^{-\tilde{V}(\xi)} &\equiv e^{ - {\rm Re}\,S(z(\xi)) - \wt(t)}, \\
  \tilde{\denom}(\xi) &\equiv \det J(\xi)\, e^{-i\, {\rm Im}\,S(z(\xi))}, 
\end{align}
we can rewrite \eqref{integral_R2} 
as a ratio of new reweighted averages, 
\begin{align}
  \langle \mathcal{O}(x) \rangle =
  \frac{\langle \tilde{\denom}(\xi)\,\mathcal{O}(z(\xi))\rangle_{\tilde\calR} }
  {\langle \tilde{\denom}(\xi) \rangle_{\tilde\calR}}, 
\end{align}
where 
\begin{align}
  \langle f(z)\rangle_{\tilde\calR} \equiv
  \frac{\int_{\tilde\calR} d \xi \, 
  e^{-\tilde{V}(\xi)}\, f(\xi)} 
  {\int_{\tilde\calR} d \xi \, 
  e^{-\tilde{V}(\xi)}}.
\label{expec_xi}
\end{align}
The weight $e^{-\wt(t)}$ is determined 
so that the function  
\begin{align}
  \tilde{Z}(t; \wt) \equiv e^{-\wt(t)} \int_{\bbR^N} dx \,
  e^{-{\rm Re}\,S(z(t,x))} 
\label{ideal_weight2}
\end{align}
is almost independent of $t$, 
as in subsection \ref{sec:weight}. 

The distribution $e^{-\tilde{V}(\xi)} / Z_{\tilde{\calR}}$ 
$(Z_{\tilde{\calR}}=\int_{\tilde{\calR}} d\xi\,e^{-\tilde{V}(\xi)})$ 
can be obtained from a Markov chain
without evaluating $J$ explicitly, 
if one uses the Metropolis-Hastings algorithm to update a configuration.%
\footnote{
  One can also use the HMC algorithm in principle, 
  but this requires the computation of the Jacobian $J=({J^i}_a(\xi))$ 
  because Hamilton's equation in molecular dynamics 
  involves the gradient 
  $\partial_{x^a} S(z(\xi))=\partial_{z^i}S(z)\,{J^i}_a(\xi)$. 
} 
Namely, from a configuration $\xi$  
we first propose a new configuration $\xi'$ 
with a probability 
\begin{align}
  {\rm Prop}(\xi'|\xi)
  \equiv \frac{1}{\sqrt{2\pi\sigma^2_t}} e^{ -(t'-t)^2/2\sigma^2_t}
  \frac{1}{(2\pi\sigma^2_x)^{N/2}} e^{ -(x'-x)^2/2\sigma^2_x}, 
\label{met_gaussian}
\end{align}
where we have treated $t$ and $x$ anisotropically. 
We then accept $\xi'$ with a probability  
\begin{align}
  \min \Bigl( 1,\,
  \frac{ {\rm Prop}(\xi|\xi')\, e^{-\tilde{V}(\xi')} }
  { {\rm Prop}(\xi'|\xi)\, e^{-\tilde{V}(\xi) } } \Bigr). 
\end{align}
This algorithm gives a stochastic process 
with the transition matrix%
\footnote{
  The diagonal components are determined 
  by the probability conservation, 
  $\int d\xi'\,\tilde{P}(\xi'|\xi)=1$.
} 
\begin{align}
  \tilde{P}(\xi'|\xi) \equiv 
  \min \Bigl( 1,\,
  \frac{ {\rm Prop}(\xi|\xi')\, e^{-\tilde{V}(\xi') } }
  { {\rm Prop}(\xi'|\xi)\, e^{-\tilde{V}(\xi) } } \Bigr)\,
  {\rm Prop}(\xi'|\xi)\quad (\xi'\neq\xi), 
\end{align}
and one can easily show that 
this satisfies the detailed balance condition:
\begin{align}
  \tilde{P}(\xi'|\xi)\,e^{-\tilde V(\xi)} = 
  \tilde{P}(\xi|\xi')\,e^{-\tilde V(\xi')}. 
\end{align}

In a generic case, 
${\rm Re}\,S(z(t,x))$ changes rapidly at large flow times $t$, 
and thus we should better change the proposal  
depending on $\xi = (t,x^a)$. 
One way is to change $(\sigma^2_t,\sigma^2_x)$ 
by randomly taking them from a set 
$\tilde{\mathcal{C}}=\{(\sigma^2_{t,c},\sigma^2_{x,c})\}$ 
$(c=1,\ldots,c_{\rm max})$ 
as in subsection \ref{sec:setup}. 
Another way is to use an asymmetric proposal ${\rm Prop}(\xi'|\xi)$ 
by making $\sigma^2_t$ and $\sigma^2_x$ $t$-dependent functions: 
\begin{align}
  {\rm Prop}(\xi'|\xi)
  = \frac{1}{\sqrt{2\pi\sigma^2_t(t)}}\, e^{-(t'-t)^2/2\sigma^2_t(t)}
  \frac{1}{(2\pi\sigma^2_x(t))^{N/2}}\, e^{-(x'-x)^2/2\sigma^2_x(t)}. 
\label{met_gaussian'}
\end{align}
In the latter case, 
the functional form of $\sigma^2_t(t)$ and $\sigma^2_x(t)$ 
are fixed manually or adaptively from test runs. 

After a sample is obtained for the region $[T_0,T_1]$, 
we consider subsamples for various estimation ranges $[\hat T_0,\hat T_1]$, 
and estimate an observable 
by looking at a plateau in the two-dimensional parameter space 
$\{(\hat T_0,\hat T_1)\}$, 
as in subsection \ref{sec:estimation}.

\baselineskip=0.9\normalbaselineskip




\begin{thebibliography}{99}

\bibitem{Aarts:2015tyj} 
G.~Aarts,
``Introductory lectures on lattice QCD at nonzero baryon number.''
J.\ Phys.\ Conf.\ Ser.\  {\bf 706}, no.\ 2, 022004 (2016)
[arXiv:1512.05145 [hep-lat]].

\bibitem{Pollet:2012} 
L.~Pollet,
``Recent developments in Quantum Monte-Carlo simulations 
with applications for cold gases,''
Rep.\ Prog.\ Phys.\  {\bf 75}, 094501 (2012)
[arXiv:1206.0781 [cond-mat.quant-gas]].

\bibitem{Parisi:1984cs} 
G.~Parisi,
``On complex probabilities,''
Phys.\ Lett.\  {\bf 131B}, 393 (1983).

\bibitem{Klauder:1983nn}
J.~R.~Klauder,
``Stochastic quantization,''
Acta Phys.\ Austriaca Suppl.\ \textbf{25}, 251-281 (1983)

\bibitem{Klauder:1983sp}
J.~R.~Klauder,
``Coherent state Langevin equations for canonical quantum systems 
with applications to the quantized Hall effect,''
Phys.\ Rev.\ A \textbf{29}, 2036-2047 (1984)

\bibitem{Aarts:2009uq}
G.~Aarts, E.~Seiler and I.~O.~Stamatescu,
``The complex Langevin method: When can it be trusted?,''
Phys.\ Rev.\ D \textbf{81}, 054508 (2010)
[arXiv:0912.3360 [hep-lat]].

\bibitem{Aarts:2011ax} 
G.~Aarts, F.~A.~James, E.~Seiler and I.~O.~Stamatescu,
``Complex Langevin: Etiology and diagnostics of its main problem,''
Eur.\ Phys.\ J.\ C {\bf 71}, 1756 (2011)
[arXiv:1101.3270 [hep-lat]].

\bibitem{Aarts:2013uxa} 
G.~Aarts, L.~Bongiovanni, E.~Seiler, D.~Sexty and I.~O.~Stamatescu,
``Controlling complex Langevin dynamics at finite density,''
Eur.\ Phys.\ J.\ A {\bf 49}, 89 (2013)
[arXiv:1303.6425 [hep-lat]].

\bibitem{Nagata:2016vkn} 
K.~Nagata, J.~Nishimura and S.~Shimasaki,
``Argument for justification of the complex Langevin method 
and the condition for correct convergence,''
Phys.\ Rev.\ D {\bf 94}, no.\ 11, 114515 (2016)
[arXiv:1606.07627 [hep-lat]].

\bibitem{Witten:2010cx}
E.~Witten,
``Analytic continuation of Chern-Simons theory,''
AMS/IP Stud.\ Adv.\ Math.\ \textbf{50}, 347-446 (2011)
[arXiv:1001.2933 [hep-th]].

\bibitem{Cristoforetti:2012su} 
M.~Cristoforetti, F.~Di Renzo and L.~Scorzato,
``New approach to the sign problem in quantum field theories: 
High density QCD on a Lefschetz thimble,''
Phys.\ Rev.\ D {\bf 86}, 074506 (2012)
[arXiv:1205.3996 [hep-lat]].

\bibitem{Cristoforetti:2013wha} 
M.~Cristoforetti, F.~Di Renzo, A.~Mukherjee and L.~Scorzato,
``Monte Carlo simulations on the Lefschetz thimble: 
Taming the sign problem,''
Phys.\ Rev.\ D {\bf 88}, no.\ 5, 051501(R) (2013)
[arXiv:1303.7204 [hep-lat]].
 
\bibitem{Fujii:2013sra} 
H.~Fujii, D.~Honda, M.~Kato, Y.~Kikukawa, S.~Komatsu and T.~Sano,
``Hybrid Monte Carlo on Lefschetz thimbles 
- A study of the residual sign problem,''
JHEP {\bf 1310}, 147 (2013)
[arXiv:1309.4371 [hep-lat]].

\bibitem{Alexandru:2015xva} 
A.~Alexandru, G.~Ba\c sar and P.~Bedaque,
``Monte Carlo algorithm for simulating fermions on Lefschetz thimbles,''
Phys.\ Rev.\ D {\bf 93}, no.\ 1, 014504 (2016)
[arXiv:1510.03258 [hep-lat]].

\bibitem{Alexandru:2015sua} 
A.~Alexandru, G.~Ba\c sar, P.~F.~Bedaque, G.~W.~Ridgway and N.~C.~Warrington,
``Sign problem and Monte Carlo calculations beyond Lefschetz thimbles,''
JHEP {\bf 1605}, 053 (2016)
[arXiv:1512.08764 [hep-lat]].

\bibitem{Fukuma:2017fjq} 
M.~Fukuma and N.~Umeda,
``Parallel tempering algorithm for integration over Lefschetz thimbles,''
PTEP {\bf 2017}, no.\ 7, 073B01 (2017)
[arXiv:1703.00861 [hep-lat]].

\bibitem{Alexandru:2017oyw} 
A.~Alexandru, G.~Ba\c sar, P.~F.~Bedaque and N.~C.~Warrington,
``Tempered transitions between thimbles,''
Phys.\ Rev.\ D {\bf 96}, no.\ 3, 034513 (2017)
[arXiv:1703.02414 [hep-lat]].

\bibitem{Alexandru:2017lqr}
A.~Alexandru, G.~Basar, P.~F.~Bedaque and G.~W.~Ridgway,
``Schwinger-Keldysh formalism on the lattice: 
A faster algorithm and its application to field theory,''
Phys. Rev. D \textbf{95}, no.11, 114501 (2017)
[arXiv:1704.06404 [hep-lat]].

\bibitem{Fukuma:2019wbv} 
M.~Fukuma, N.~Matsumoto and N.~Umeda,
``Applying the tempered Lefschetz thimble method 
to the Hubbard model away from half filling,''
Phys.\ Rev.\ D {\bf 100}, no.\ 11, 114510 (2019)
[arXiv:1906.04243 [cond-mat.str-el]].

\bibitem{Fukuma:2019uot}
M.~Fukuma, N.~Matsumoto and N.~Umeda,
``Implementation of the HMC algorithm 
on the tempered Lefschetz thimble method,''
[arXiv:1912.13303 [hep-lat]].
 
\bibitem{Alexandru:2019} 
A.~Alexandru, 
``Improved algorithms for generalized thimble method,''
talk at the 37th international conference on lattice field theory, Wuhan, 2019. 

\bibitem{Swendsen1986}
R.~H.~Swendsen and J.-S.~Wang,
``Replica Monte Carlo simulation of spin-glasses,''
Phys.\ Rev.\ Lett.\ {\bf 57} 2607 (1986). 
  
\bibitem{Geyer1991}
C.~J.~Geyer, 
``Markov chain Monte Carlo maximum likelihood,''
in computing science and statistics: 
\textit{Proceedings of the 23rd Symposium on the Interface}, 
American Statistical Association, New York, p.~156 (1991).

\bibitem{Hukushima1996}
K.~Hukushima and K.~Nemoto, 
``Exchange Monte Carlo method 
and application to spin glass simulations,''
J.\ Phys.\ Soc.\ Jpn.\ {\bf 65}, 1604 (1996).
 
\bibitem{Fukuma:2018qgv} 
M.~Fukuma, N.~Matsumoto and N.~Umeda,
``Emergence of AdS geometry in the simulated tempering algorithm,''
JHEP {\bf 1811}, 060 (2018)
[arXiv:1806.10915 [hep-th]].

\bibitem{Fukuma:2020got}
M.~Fukuma, N.~Matsumoto and N.~Umeda,
``Distance between configurations in MCMC simulations 
and the geometrical optimization of the tempering algorithms,''
PoS \textbf{LATTICE2019}, 168 (2019)
[arXiv:2001.02028 [hep-lat]].

\bibitem{Fukuma:2017wzs} 
M.~Fukuma, N.~Matsumoto and N.~Umeda,
``Distance between configurations in Markov chain Monte Carlo simulations,''
JHEP {\bf 1712}, 001 (2017)
[arXiv:1705.06097 [hep-lat]].
  
\bibitem{Stephanov:1996ki} 
M.~A.~Stephanov,
``Random matrix model of QCD at finite density 
and the nature of the quenched limit,''
Phys.\ Rev.\ Lett.\  {\bf 76}, 4472 (1996)
[hep-lat/9604003].

\bibitem{Halasz:1998qr} 
M.~A.~Halasz, A.~D.~Jackson, R.~E.~Shrock, M.~A.~Stephanov and 
J.~J.~M.~Verbaarschot,
``On the phase diagram of QCD,''
Phys.\ Rev.\ D {\bf 58}, 096007 (1998)
[hep-ph/9804290].
   
\bibitem{Bloch:2017sex}
J.~Bloch, J.~Glesaaen, J.~J.~M.~Verbaarschot and S.~Zafeiropoulos,
``Complex Langevin Simulation of a Random Matrix Model 
at Nonzero Chemical Potential,''
JHEP \textbf{03}, 015 (2018)
[arXiv:1712.07514 [hep-lat]].

\bibitem{Andersen:1983}
H.~C.~Andersen,
``RATTLE: A ``velocity'' version of the SHAKE algorithm 
for molecular dynamics calculations,''
J.\ Comput.\ Phys.\ {\bf 52}, 24 (1983).

\bibitem{Leimkuhler:1994} 
B.~J.~Leimkuhler and R.~D.~Skeel, 
``Symplectic numerical integrators in constrained Hamiltonian systems,''
J.\ Comput.\ Phys.\ {\bf 112}, 117 (1994).

\bibitem{Saad:1983}
Y.~Saad and M.~H.~Schultz,
``GMRES: A Generalized minimal residual algorithm 
for solving nonsymmetric linear systems,''
SIAM J.\ Sci.\ and Stat.\ Comput.\ {\bf 7}, 856 (1986).
   
\bibitem{Vorst:1990}
H.~A.~van der Vorst,
``Bi-CGSTAB: A fast and smoothly converging variant of Bi-CG 
for the solution of nonsymmetric linear systems,''
SIAM J.\ Sci.\ and Stat.\ Comput.\ {\bf 13}, 631 (1992).
  
\bibitem{Wang:2000fzi}
F.~Wang and D.~P.~Landau,
``Efficient, multiple-range random walk algorithm 
to calculate the density of states,''
Phys.\ Rev.\ Lett.\ \textbf{86}, no.10, 2050 (2001)
[arXiv:cond-mat/0011174 [cond-mat.stat-mech]].

\bibitem{Marinari:1992qd} 
E.~Marinari and G.~Parisi,
``Simulated tempering: A new Monte Carlo scheme,''
Europhys.\ Lett.\ \textbf{19}, 451-458 (1992)
[hep-lat/9205018].

\bibitem{Tanizaki:2017yow}
Y.~Tanizaki, H.~Nishimura and J.~J.~M.~Verbaarschot,
``Gradient flows without blow-up for Lefschetz thimbles,''
JHEP \textbf{10}, 100 (2017)
[arXiv:1706.03822 [hep-lat]].

\bibitem{FMU_JPS2020} 
M.~Fukuma, N.~Matsumoto and N.~Umeda, 
``Applying the tempered Lefschetz thimble method 
to the sign problem of chiral matrix models 
and the estimation of the computational scaling,'' 
talk at the JPS meeting (online), September 15, 2020.

\end{thebibliography}
\end{document}